\documentclass[twocolumn]{aastex631}
\usepackage{amsmath}
\usepackage{amssymb}
\usepackage{multirow}
\usepackage{tablefootnote}
\usepackage[T5]{fontenc}

\newcommand{\fesc}{\ifmmode{f_{\rm esc}}\else{$f_{\rm esc}$}\fi}
\newcommand{\fescs}{\ifmmode{f_{\rm esc}^\star}\else{$f_{\rm esc}^\star$}\fi}
\newcommand{\kms}{\ifmmode{{\;\rm km~s^{-1}}}\else{km~s$^{-1}$}\fi}
\newcommand{\fgas}{\ifmmode{{f_{\rm gas}}}\else{$f_{\rm gas}$}\fi}
\newcommand{\cubecm}{\ifmmode{{\rm cm^{-3}}}\else{cm$^{-3}$}\fi}
\newcommand{\ztwo}{\ifmmode{{\rm [Z_2/H]}}\else{[Z$_2$/H]}\fi}
\newcommand{\zthree}{\ifmmode{{\rm [Z_3/H]}}\else{[Z$_3$/H]}\fi}
\newcommand{\lsim}{\lower0.3em\hbox{$\,\buildrel <\over\sim\,$}}
\newcommand{\gsim}{\lower0.3em\hbox{$\,\buildrel >\over\sim\,$}}

\newcommand{\sfr}{\ifmmode{\textrm{M}_\odot \,\textrm{yr}^{-1} \,\textrm{Mpc}^{-3}}\else{M$_\odot$ yr$^{-1}$ Mpc$^{-3}$}\fi}
\newcommand{\hsfr}{\ifmmode{\textrm{M}_\odot\, \textrm{yr}^{-1}}\else{M$_\odot$ yr$^{-1}$}\fi}

\newcommand{\eavg}{\ifmmode{\langle E_\gamma \rangle}\else{$\langle E_\gamma \rangle$}\fi}

\newcommand{\Ms}{\ifmmode{M_\odot}\else{$M_\odot$}\fi}
\newcommand{\vrms}{\ifmmode{v_{\rm rms}}\else{$v_{\rm rms}$}\fi}

\newcommand{\tvir}{\ifmmode{T_{\rm{vir}}}\else{$T_{\rm{vir}}$}\fi}
\newcommand{\mvir}{\ifmmode{M_{\rm{vir}}}\else{$M_{\rm{vir}}$}\fi}
\newcommand{\rvir}{\ifmmode{r_{\rm{vir}}}\else{$r_{\rm{vir}}$}\fi}

\newcommand{\jj}{\ifmmode{J_{21}}\else{$J_{21}$}\fi}
\newcommand{\flw}{\ifmmode{F_{LW}}\else{$F_{LW}$}\fi}
\newcommand{\kph}{\ifmmode{k_{\rm ph}}\else{$k_{\rm ph}$}\fi}

\newcommand{\zsun}{\ifmmode{\rm\,Z_\odot}\else{$\rm\,Z_\odot$}\fi}

\newcommand{\nhi}{\ifmmode{N_{\rm HI}}\else{$N_{\rm HI}$}\fi}

\begin{document}

\title{Solving Milky Way-sized Systems with {\sc Haskap Pie}: A Halo finding Algorithm with efficient Sampling, K-means clustering, tree-Assembly, Particle tracking, Python modules, Inter-code applicability, and Energy solving}

\author[0000-0002-8638-1697]{Kirk S.~S.~Barrow}
\affiliation{Department of Astronomy, University of Illinois at Urbana-Champaign, 
1002 W Green St, Urbana, IL 61801, USA}

\author[0009-0002-2290-8039]{Thịnh Hữu Nguyễn}
\affiliation{Department of Astronomy, University of Illinois at Urbana-Champaign, 
1002 W Green St, Urbana, IL 61801, USA}
\affiliation{Center for AstroPhysical Surveys, National Center for Supercomputing Applications, Urbana, IL 61801, USA}

\author[0009-0005-5945-209X]{Edward Skrabacz}
\affiliation{Department of Astronomy, University of Illinois at Urbana-Champaign, 
1002 W Green St, Urbana, IL 61801, USA}

\begin{abstract}

We describe a new Python-based stand-alone halo finding algorithm, {\sc Haskap Pie}, that combines several methods of halo finding and tracking into a single calculation. Our halo-finder flexibly solves halos for simulations produced by eight simulation codes (ART-I, ENZO, RAMSES, CHANGA, GADGET-
3, GEAR, AREPO, and GIZMO) and for both zoom-in or full-box N-body or hydrodynamical simulations and includes a unified, robust set of pre-tuned parameters. When compared to \textsc{Rockstar} \citet{Behroozi_2013} and {\sc Consistent Trees} \citep{2013ApJ...763...18B}, our halo-finder tracks subhalos much longer and more consistently, produces halos with better constrained physical parameters, and returns a much denser halo mass function for halos with more than 100 particles. Our results also compare favorably to recently described specialized particle-tracking extensions to \textsc{Rockstar}. {\sc Haskap Pie} is well-suited to a variety of studies of simulated galaxies and is particularly robust for a new generation of studies of merging and satellite galaxies. For our initial paper, we focus on describing our algorithm's ability to find and track halos and subhalos in complex Milky Way-sized halo systems.

\end{abstract}

\section{Introduction}
Cosmological simulations have proven to be vital to our understanding of dark matter, galaxy assembly, astrophysical processes, Reionization, and the constraints on the fundamental cosmological parameters. 

Any study of galaxies in large-scale simulations requires an algorithm that identifies and parametrizes the presence of dark matter halos from amongst the cosmic web: a halo-finder. Though their existence is well-described in spherical collapse theory, the definition of a dark matter halo largely depends on the manner in which they are calculated. Several studies have found that halo-finding and halo-tree algorithms can return different results based on the techniques and routines applied \citep[e.g.][]{2011MNRAS.415.2293K,2012MNRAS.423.1200O,2013MNRAS.436..150S}. As layers of more sophisticated methods for the analysis of simulations proliferate through the community, the viability of the basic simulation metadata about the presence of halos and galaxies needs to be scrutinized, as inconsistencies and inaccuracies can have compounding effects on scientific predictions throughout the field. 

These halo-finders are generally classified by their algorithmic strategies and currently fall into five general categories. Friends-of-Friends (FoF) halo-finders \citep[e.g.][]{1985ApJ...292..371D,2001MNRAS.328..726S,2009MNRAS.399..497D,2010ApJS..191...43S} work by associating dark matter particles that are separated by a maximum linking length with each other, creating a chain of associations that roughly enclose an overdense region. Algorithms of this nature are typically efficient and scalable to very large simulations. However, FoF halo-finders can miss smaller halos and subhalos in a large complex of halos as this method does not determine whether individual particles are gravitationally bound.

Spherical overdensity-based halo finders \citep[e.g.][]{1994MNRAS.271..676L,1998ApJ...498..137E,2009ApJS..182..608K,2010A&A...519A..94P,2022MNRAS.509..501H,2022A&A...664A..42V} address this by associating particles with a density field and applying one of many methods, such as interpolating an adaptive mesh refinement grid of densities, to deduce the presence of halos from the particle density fields. These finders reduce incidences of miss-associated particles and improve the identification of subhalos, while still being computationally efficient. However, they also do not directly calculate whether particles are bound to halos when halos are initially identified and therefore are still likely to miss-define halos, miss subhalos, and miss-associate particles, for example, as found by \citet{2025MNRAS.539..776C}. To account for this, spherical overdensity-finding can be paired with unbinding procedures to improve solutions. This may include methods such as assuming a spherical potential about the overdense regions and removing particles that are not bound to an enclosed spherical mass or grouping particles by their velocity dispersions.

A phase-space halo finder \citep[e.g.][]{2006ApJ...649....1D,2009MNRAS.396.1329M,2019PASA...36...21E} builds on an FoF to associate groups of particles spatially. The phase-space halo-finder {\sc Rockstar} \citep{Behroozi_2013}, which quickly grew to become a standard halo-finder in the community, incorporates positional and velocity dispersions to adaptively determine linking lengths. Using this phase-space has the advantage of operating calculations on the appropriate scale of subhalos within larger halos and leads to a much more complete list of candidates within complex substructures. Candidate halos are then connected hierarchically and, using a Barnes-Hut algorithm, unbound particles are removed. The Barnes-Hut algorithm  estimates whether particles are bound using an octree procedure, which is similar to methods employed by \citet{1999ApJ...516..530K} and \citet{2011arXiv1109.0003R}. While phase-space halo-finding is largely successful at identifying almost all likely halos, pruning this list to include only physically bound structures or recovering consistent results across timesteps is not always trivial.

Finally, temporal information can be used to improve the permanence of viable halos across time with a halo-tracking algorithm. This class includes dedicated halo tree codes such as {\sc Consistent Trees} \citep{2013ApJ...763...18B}, which builds on halo metadata to construct halo trees that can extrapolate and interpolate halo properties over time to recover and connect missing halos. Dedicated particle tracking algorithms, such as {\sc SYMFIND} \citep{2024ApJ...970..178M} or the algorithms by by \citet{2024MNRAS.533.3811D} and \citet{2025arXiv250310766K}, often use outputs from {\sc Consistent Trees} or from FoF codes such as {\sc HBT/HBT+} \citep{2012MNRAS.427.2437H,2018MNRAS.474..604H}, which both have their own temporal tracking methods, to further refine the temporal tracking of halos. These methods have been shown to improve the tracking of partially disrupted subhalos using particle tracking when they would otherwise be inscrutable with other halo-finding methods. However, particle tracking methods that use halo finders as a baseline and then reanalyze or extend their data products are, by construction, limited to the halos discovered by their baseline halo-finding methods.

More robust determinations of gravitational boundedness, such as methods that both identify and track halos by conducting searches for self-bound groups of particles, can produce well-defined halos at the expense of higher computational cost. For instance, directly calculating the potential energy of a system of particles scales poorly with the number of particles, $n$. Calculating the potential energy in this way means calculating the distance between each pair of particles, typically iteratively, until the correct center of mass can be found. Using efficient vectorized algorithms to simulaneously batch particles and solve for their distances can speed up calculations. However, holding even $n^2$ relative positions in memory becomes prohibitive after just a few million particles. The memory and computational cost both necessitate {\sc Rockstar}'s use of an estimate for their version of this calculation.

Motivated by the needs of studies of galaxy-merging complexes, we examined whether adding an additional energy solving or particle tracking method to results from {\sc Rockstar} would satisfy our modeling needs. In this study, we found that there was the potential to make much more progress in finding and tracking many more halos if we built our new algorithm, {\sc Haskap Pie}. In Sec. \ref{sec: methods}, we describe our simulation data sets and detail {\sc Haskap Pie}'s techniques and methodological choices. In Sec. \ref{sec: results}, we show how {\sc Haskap Pie} provides a much more complete picture of halos, subhalos, and halo dynamics than the combination of {\sc Rockstar} and {\sc Consistent Trees} (hereafter abbreviated as `RCT'). Sec. \ref{sec: results} also compares recent particle-tracking extensions before we summarize our findings in Sec. \ref{sec: summary}.

\section{Halo-Solving Methods}
\label{sec: methods}

We express a preference for a definition of ``dark matter halos" wherein they are gravitationally self-bound and exist persistently with at least some of the same dark matter composition bound to the halo over time. This contrasts with halos defined by either a fixed overdensity, solely by self-gravitation wherein components are free to join or leave the region, particle position and velocity dispersions (phase space), or halos defined by particle linking lengths (FoF).

However, a combination of each of these definitions finds utility in our analysis to address edge cases such as complex subhalo populations that cannot be tracked with a rigid definition of halos. Our halo-finding method includes overdensity-finding, particle cluster determination, energy-solving, particle-tracking (forward and backward propagation), and data reduction with several iterative steps. This combination of methods ensures that halos are recovered more consistently.

The default parameters used during our testing are explained in this description of our method, and these parameters were found to be appropriate for our diverse sample of hydrodynamical cosmological and zoom-in simulations, but several of these parameters could potentially be further tuned or improved upon for user circumstances to produce more relevant results. In Sec. \ref{sec: parameters}, we have included a detailed description of how we have selected our modeling parameters and how these choices affect the final halo-finding results.

We have yet to test our technique on truly massive simulations with hundreds of thousands of halos or hundreds of millions of particles, which may require further tuning especially in the overdensity-finding routine, which is tuned to be complete for zoom-in and full-box simulations with a 512$^3$ root grid. For those simulations, we would also likely need to explore further optimization strategies. Our intention is to converge on a single set of parameters for all use cases, and these will be reported in future analyses as development continues.

\subsection{Simulations Analyzed}
\label{sec: sims}

We perform our halo-finding on N-body and radiative-hydrodynamic zoom-in simulations. Our primary N-body simulation was run using ENZO \citep{Bryan2014} by initializing a 512$^3$ root grid and a (5 Mpc/h)$^3$ box initialized with a flat $\Lambda$CDM cosmology and are run with the cosmological parameters taken from \citet{PlanckCollaboration+2016}: $\Omega_{M} = 0.3111$, $\Omega_{\Lambda} = 0.6889$, $h = 0.6766$, $\sigma_{8} = 0.8102$, and $n = 0.9665$, resulting in a dark matter resolution of $\sim1.2 \times 10^5$ M$_\odot$.

The AGORA collaboration recently published their analysis on a zoom-in region run with common initial conditions and prescriptions with star formation, feedback, and radiative transfer across eight cosmological simulation codes to at least $z=2$ \citep[][\texttt{Cosmorun-2}]{2024ApJ...968..125R} (ART-I, ENZO, RAMSES, CHANGA, GADGET-
3, GEAR, GIZMO, and AREPO) with an effective dark matter mass resolution of $\sim2.8 \times 10^5$ M$_\odot$. We have solved halo trees for each of these codes and present them as part of our analysis. The data analysis code \texttt{yt} \citep{2011ApJS..192....9T} is used to process, analyze, and plot data from all simulations. For AGORA's GEAR and GIZMO data, we resave the particle data to bypass the \texttt{yt} interface for particle reading, which can return incomplete particle lists for these data. For AGORA's CHANGA data, we both resave the particle data and reassign the particle IDs to facilitate faster loading times and bypass \texttt{yt}'s ID assignment, which can be inconsistent for these data. We pay particular attention to the results from ART-I for our detailed analyses as simulation timesteps to $z=0$ were available at the time of writing, halo trees were easily and quickly solved, and \texttt{yt} returned complete simulation data without a workaround. 

%For the examples in our description of the methods, we use the ENZO version of Cosmorun-2.

For the examples in our description of our energy-solving technique in Sec. \ref{sec: energy}, we also use an ENZO high-redshift radiative-hydrodynamic zoom-in simulation described in \citet{2024ApJ...969..144S} with an effective dark matter mass resolution of $2.81 \times 10^4$ M$_\odot$ centered around a $\sim 1.3 \times 10^9$ M$_\odot$ halo at z=7.5.

\subsubsection{Refined Regions}
\label{sec:Refined Regions}
Zoom-in simulations typically surround a high-resolution volume with less well-defined halos made of unrefined dark-matter particles. We have included an optional algorithm that automatically detects a region that consists of refined dark matter particles, which may be a subset of the simulation-defined refined region as massive particles can migrate within the intended refined region as the simulation progresses. When using this algorithm, halo finding will populate a tree only from within the uncontaminated volume when our region-finding algorithm is called, which can aid the analyses of physical phenomena in zoom-in simulations. In this study, we define an uncontaminated refined region as a region containing only the most refined and the second most refined dark matter particles (particles at the lowest and second lowest mass level, respectively). A dark matter particle whose mass is larger than the second most refined mass level is classified as an unrefined particle. 

The process of determining the uncontaminated refined region begins with loading dark matter particles in the whole simulation box to identify all the available dark matter mass levels. Occasionally, there can be spurious dark matter masses present in simulation data, such as masses that are very small or masses that are a few units off from the actual dark matter mass levels. To create a consistent list of the dark matter mass levels, we require that the number of dark matter particles of a certain mass needs to be at least 0.5\% the total number of dark matter particles in the whole box. This is to correct for small errors in particle masses that may accumulate during simulations that may produce particles with unique masses. Next, with the particle positions and the list of all mass levels, we locate the center of mass of all the most refined particles. From this center, we iteratively expand out in each of the six directions (two directions for each of the three axes) in the box. The length of each expansion step starts at 1/160 of the simulation box's size. If an unrefined particle is found in one expanding direction, we reduce the step size in that direction by 1.5 times and re-check the existence of unrefined particles. This iterative change in step size allows more efficient computation. The expansion in a direction is stopped when the step size of that direction is less than 1/10000 of the simulation box's size. When the expansion is stopped in all six directions, a refined region is found.

We have also included a separate algorithm to quickly and automatically determine the presence of dark matter particles with different masses, which would imply a refined region, before invoking our refined region-finding algorithm. If the sampled particles have the same mass, we skip this procedure and run our halo-finding on the full simulated volume. This eliminates any need for users to post-process halo trees or input parameters for zoom-in versus non-zoom-in simulations. The current version of our halo-finding code only requests the name of the simulation code, a path to the simulation, and a save path as input parameters. There is also an optional parameter for skipping timesteps for simulations with more outputs than are needed for halo-tree calculations.

\subsection{Algorithm Steps}

\label{sec: steps}

The quality and demographics of our halo results are potentially sensitive to the order of and repetition of the steps of our halo-finding algorithm. Therefore, we build redundancies into the process to ensure results are well-recovered for our use cases. For a simulation where data are saved as snapshots (timesteps) corresponding to an ordered sequence of times after the Big Bang, the algorithm starts at the latest snapshot and proceeds backward in time with short periods of moving forward in time. Once reaching the first snapshot, the algorithm moves forward in time until the final snapshot is reached again, which means each timestep is analyzed with forward and backward modelling at least once. During testing and development, the following configuration was found to be effective for simulations that span the Hubble Time or shorter:

\begin{enumerate}
    \item \textbf{Last Snapshot:} Overdensity-Finding (FoF) / Particle Cluster-Finding 
    \item \textbf{Prior 1/11th of Snapshots:} Backward-Modeling / Particle Cluster-Finding
    \item \textbf{Prior Snapshot:} Overdensity-Finding (FoF) / Backward-Modeling / Particle Cluster-Finding
    \item \textbf{Prior Two Snapshots:} Backward-Modeling / Particle Cluster-Finding
    \item \textbf{Next 1/25th of Snapshots:} Forward-Modeling / Particle Cluster-Finding
    \item Delete halos that have short histories and a minimum timestep $>$ last overdensity-finding step
    \item \textbf{From Earliest Unmodeled Snapshot:} Repeat last five steps until reaching the first timestep
    \item \textbf{First to Last Snapshot:} Forward-Modeling / Particle Cluster Finding
    \item Clean and prune the final halo tree
\end{enumerate}

To better explain the steps, here is an example for a simulation with 275 snapshots. We start the halo finder with Step 1 at Snapshot 275. Then, Step 2 runs from Snapshot 274 to Snapshot 250. Next, Step 3 is for Snapshot 249, and Step 4 is for Snapshots 248 and 247. In the forward modeling in Step 5, the code is launched from Snapshot 247 to 257. The whole cycle will begin again at Step 7 by covering Snapshot 246 to Snapshot 218. Similarly, subsequent cycles will proceed until Snapshot 1 is reached. Lastly, Step 8 will run from Snapshot 1 to Snapshot 275.

Note that if the simulation has fewer than 25 saved snapshots, step 5 is skipped. If there are fewer than 11 snapshots, step 3 begins at the third to last snapshot, and steps 5-7 are skipped. At least five simulation snapshots are required to run all the intended routines of this algorithm, which are described in more detail in the following sections. For a simulation that spans the Hubble time, the application of our algorithmics is suggested only if there are a minimum of at least 100 timesteps to see the best tracking results to keep in line with our test suite and use cases.

As discussed in Sec. \ref{sec: parameters}, the overdensity-finding step tends to overpopulate our halo-catalog and so we do not need to perform it for every timestep. However, rather than find a minimum number of times we can run that algorithm without jeopardizing our solution, we choose run it more times than necessary and prune the result to create our final halo catalogs. Therefore, this algorithm represents a setup for which we know that our choices do not limit the solution. We find that for a cosmological simulation spanning the Hubble time, nearly complete halo lists and trees can be generated by invoking this process no more than eleven times for $z<6$.

\subsection{Overdensity Finding}

The identification of candidate halos begins at the final (latest) timestep of the simulation with an overdensity finder. We split the simulation volume into several scales of sub-volumes. Modules included with \texttt{yt} efficiently produce ``deposited" mass fields from enclosed dark matter particles on an arbitrary grid and interface with several simulation codes. Beginning with the full volume, we use \texttt{yt} to generate a $30^3$ grid of dark matter overdensities using the calculated simulation critical density. Adjacent grid cells with overdensities greater than forty times the critical density of the universe, $\rho_c$ = $\frac{3 H(z)^2}{8 \pi G}$, are connected (similar to using a friends-of-friends algorithm) to produce coarse overdense volumes. The threshold of 40 identifies which sub-voumes of the simulation should contain small halos. A simulation with resolution less than 103$^3$ and an overdensity greater than 300 would return an overdensity of at least 40 in a $30^3$ search. That process is repeated for $90^3$ and $270^3$ grids which allows us to handle simulations or zoom-in regions with resolutions up to 512$^3$. Within these regions, we iteratively search with finer grids ($(90N)^3$, where $N$ goes from 3 to 9) gradually until we find all regions with overdensities greater than 300$\rho_c$. Depending on the resolution of the simulation, the smallest halos may not be recognized by the over-density finding procedure. Further development of this will accompany our optimization of our aglorithm for larger simulations and volumes in future work.

\begin{figure*}
\begin{center}
\includegraphics[width=\linewidth]{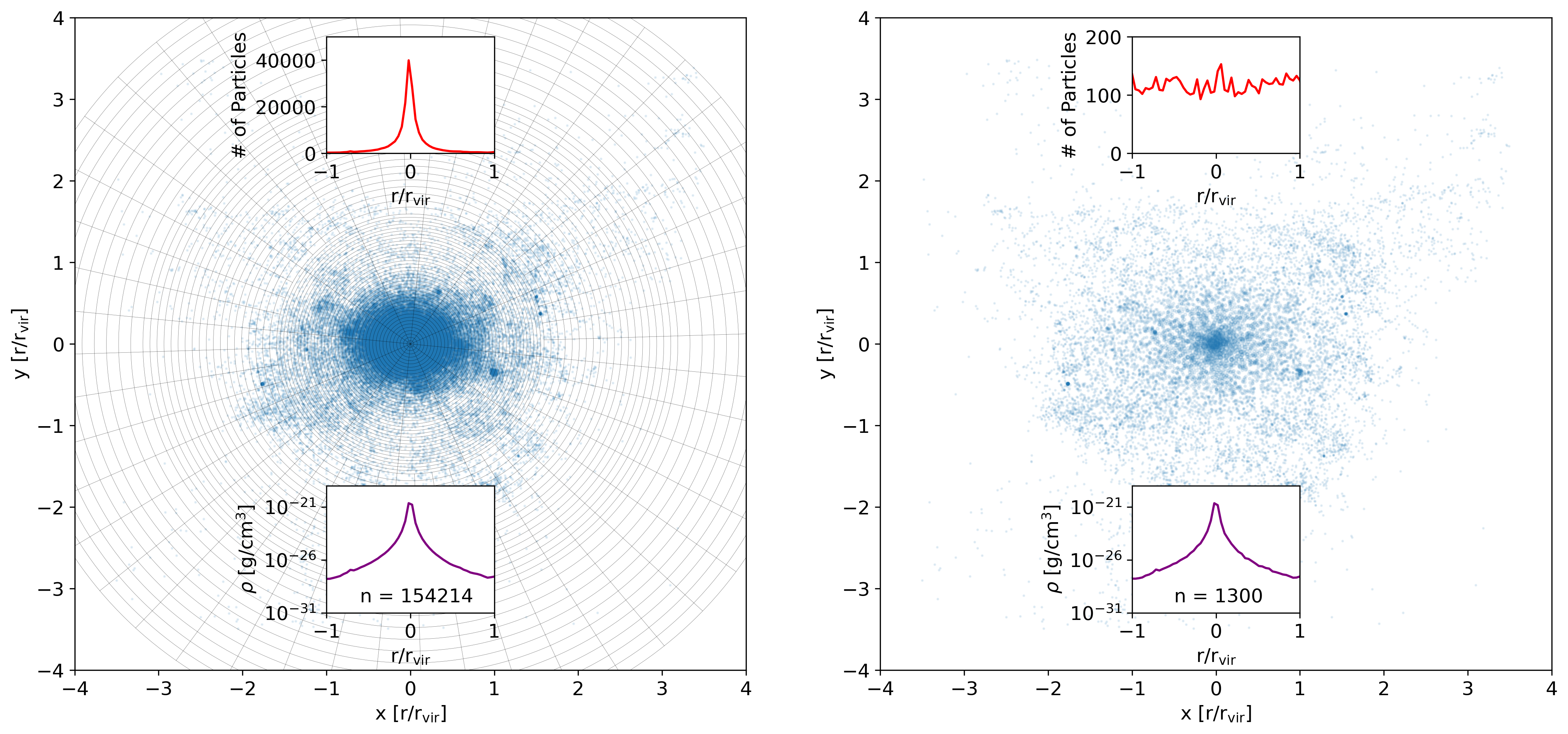}
\caption{2-D demonstration of the particle sampling technique. Left is a 2-D slice of the particles about the main halo in the AGORA ENZO simulation at $z= 0.0989$ and right is the region after sampling in 12 (inner region) or 48 (outer regions) directions in the $x-y$ plane at 70 radii, which are lightly shaded into the left plot. The size of the points is proportional to the square root of their mass. Insets show the density (purple) and particle number distributions (red) as labeled from the center by taking a 1-D slice in the $x$-direction. The bottom inset in each plot reports the remaining number of particles after both slices of the original 7,820,075 particles within a box bounding 3.5$r_{\rm vir}$. For this example, sampled densities at the center of annular sectors have a mean error of $\sim$4.1\% and no error in enclosed mass at the annuli radial boundaries despite using less than 1/110th of the particles.}
\label{fig: cuts}
\end{center}
\end{figure*}

This method entirely ignores the number of particles within the grid, and the \texttt{yt} particle depositions to the grid were efficient for all simulations tested. The determination of overdense regions in each subvolume is functionally independent and is therefore assigned to a multi-threaded task, allowing many thousands of overdense regions in a large simulation to be identified on a laptop in a few minutes or less. The longest task is typically parsing the simulation data hierarchy within the subvolumes.  

If we limit our parallelized search to smaller volumes that contain massive halos in the simulation, though we buffer the search regions, this hierarchical approach may miss main halos when pieces of a large halo are assigned to different cores, so as a final step we search the entire volume at higher resolution (270$^3$) for large regions with overdensities over 300 on a single core. This returns solutions for a large dominating overdense volume that occupies a significant fraction of a simulation, such as for a halo-centered refined region. To avoid excessive overlapping of resulting overdense volumes, the less massive of two resulting overlapping volumes is erased as redundant if the distance between their centers is less than half their half-width and the ratio of their masses or half-width is between 0.75 and 1.25.

Note that this method is only applied to the entire simulation volume at most 11 times with our solving strategy (see Sec. \ref{sec: steps}), and our other techniques, of which some also identify new halos, are employed to ensure that the halo trees are complete. While results converged with our approach for most of cosmic time, we noticed that our initial results at high redshift ($z >7$) seemed to be less complete than expected for halos falling into the largest halos in our initial tests. This is because halos coalesced and fell into main halos so quickly that they were not captured in our intermittent overdensity-finding calculations (see Sec. \ref{sec: n-body} for a discussion of collapse times). For these results, we ran our overdensity-finding algorithm more often than described in Sec. \ref{sec: steps} for volumes focused around just the twenty largest halos when $z>6$ during backward-modeling which solved the issue for all the simulations we analyzed. Specifically, we ran our overdensity-finding algorithm every $\min(7, N_{tot}/22)$ timesteps as long as Step 3 is not within the first or last $N_{tot}/55$ timesteps of the simulation, where $N_{tot}$ is the total number of timesteps. The current version of the code does this for all halos at $z>6$ over $10^9$ M$_\odot$ as well as the top twenty halos to cover more use cases.

\subsection{Particle Cluster Finding}

The volumes identified in the overdensity finder contain particles that may exist in halos, but these volumes are coarse and subhalos, mergers, and other edge cases are difficult to extract from overdensity alone. Therefore, we further refine our candidate regions using a cluster-finding routine based on N-body (dark matter) particles.

However, holding all the relevant particle data in computer memory can become a significant limiting bottleneck. Also, given our insistence on prioritizing solving for gravitational boundedness for halos that could contain tens of millions of particles, we were forced to develop a method that scales better than the $O(n_o^2)$ scaling of brute force potential energy calculations, where $n_o$ is the raw number of N-body particles from the simulation, which could easily become infeasible.

\subsubsection{Particle Sampling}
\label{sec: sampling}

We search for halos in spherical search volume either defined by our over-density-finding routine or our particle tracking routines described in Sec. \ref{sec: tracking}. Often these volume contain more particles than we need full characterize a gravitational potential well and identify halos.  If the volume contains more than 10,000 particles, from the center to one-third the radius of this volume, we create 40 equal-depth spherical shells each split into 12 HEALPix\citep{2005ApJ...622..759G}-based annular sectors (where an annulus is a spherical shell and an annular sector is a segment of a shell between an opening angle). The next third of the radius is split into 20 annuli for 48 sectors and the final third into 12 annuli for 48 sectors for a total of 2,016 annular sectors. Within each annular sector, if there are more than a set minimum number of particles, $n_{\rm min}$, then $n_{\rm min}$ particles are randomly chosen and their mass is upscaled such that these particles represent the total mass in the annular sector. The number and density of annular sectors are chosen so that the potential well of the target halo is well sampled in key regions. Specifically, the dense inner regions that define most of the potential well and the region near the halo boundary (the middle 960 annuli) where particles may represent smaller infalling halos and sub-halos. The outer 576 annuli are sensitive to the detection of nearby halos but this is a low priority and is kept intentionally coarse as it mostly serves to make sure the potential well is well-described and bound particles outside the halo radius are included.

This is demonstrated for a 2-D example in Fig. \ref{fig: cuts}. By using more massive particles, we flatten the radial number density distribution of particles (top insets) while preserving the radial mass density distribution (bottom insets) and thus the gravitational potential of the halo. In our 2-D example from Fig. \ref{fig: cuts}, for radial density bins the local density distribution had a mean absolute error fraction of 0.04052 with a standard deviation of 0.05147 as compared to the unsampled particles. When using the full 3-D procedure, there is no error in the cumulative mass within the halo radius or at any of the 70 shell radii by construction. Therefore, any errors resulting from our sampling procedure can only result in small, localized deviations to the shape of the potential, but not to the enclosed mass at each annular sector. In a spherically symmetric density profile, this would converge to the exact answer for gravitational attraction at each annular sector. For non-spherical density profiles, our method allows us to constrain the perturbations by increasing the sampling density.

The results reported for {\sc Haskap Pie} are based on using  $n_{\rm min}$ = max(10, 100$r_{\rm shell}/r_{o}$) for each annular sector, where $r_{\rm shell}$ is the radius to the outer edge of the annulus, and $r_o$ is the radius of the spherical search volume. This scaling ensures that wider annular sectors have a correspondingly higher particle representation. Since we use more HEALPix directions at larger radii, the scaling compounds, resulting in far more particles per radian at farther radii than in the center of the halo. The minimum value of 10 means that halo regions with fewer than 20,160 particles must contain annular sectors with their full particle lists and small halos have every or almost every constituent particle represented. Because each halo is represented by particles local to the 2,016 annular sectors, our sampling also provides a good spatial representation of the particles in the original halo and halo definitions are well-converged to results using the full particle list, leading to stable definitions of halo centers.

\begin{figure*}[t!]
\begin{center}
\includegraphics[width=\linewidth]{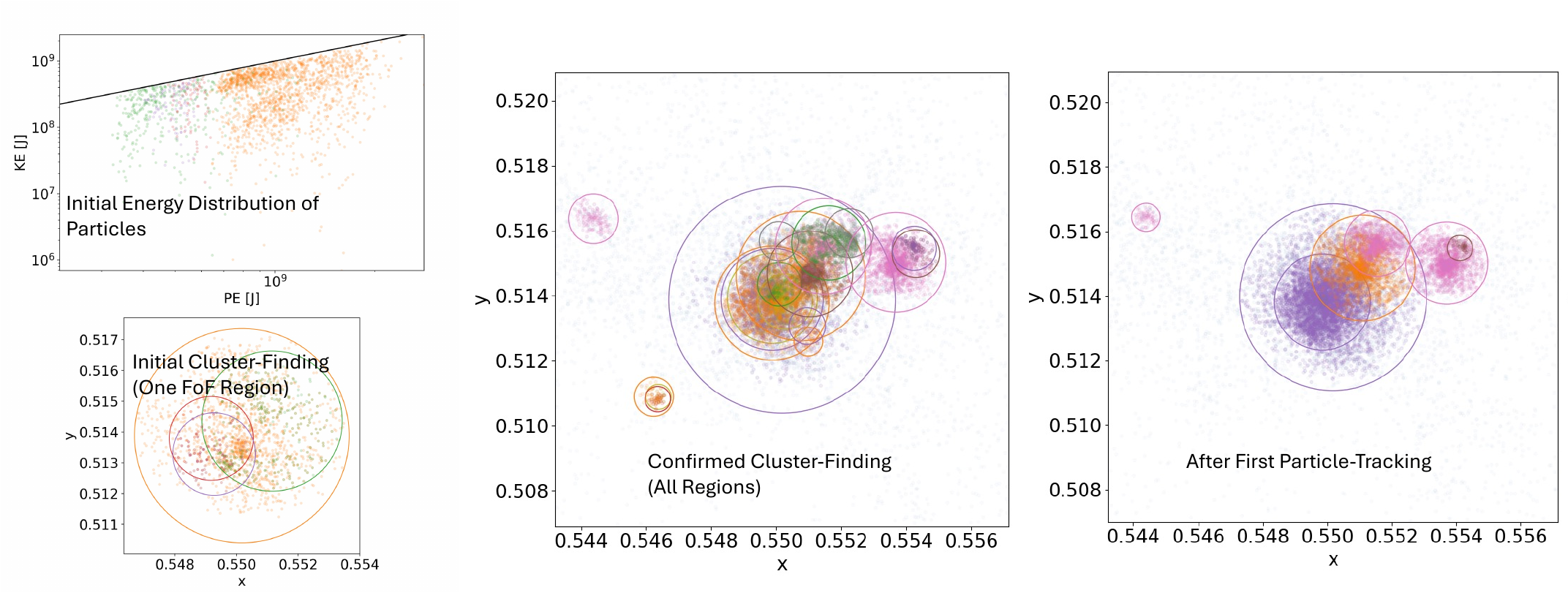}
\caption{Figures showing the evolution of halo solving through our pipeline.  Left: Clusters of halos and subhalos found for one candidate overdense volume with three iterations of k-means clustering showing the energy distributions of the halos (top) and their physical extent (bottom). Colored scatter points are particles found to be bound to the halos. Center: Overlapping results of all clusters found for all overdense volumes. Right: Halos confirmed by backward modeling and pruning for one timestep.  Our method overpopulates halos and subhalos and then prunes by only including halos that have a sustained, unique physical presence across at least five consecutive timesteps. The visually apparent flattening of the particle number density distribution aids the identification of subhalos using k-means clustering.}
\label{fig:method 1}
\end{center}
\end{figure*}

The result of the procedure is that the regions local to halos with extremely high particle counts such as the $10^{12}$ M$_\odot$ halos (for example 7,820,075 particles within 3.5$r_{\rm vir}$) in a late timestep of the AGORA ENZO simulation (see Sec. \ref{sec: sims}) can be represented with at most 104,643 particles and in practice were represented by about 62,700 particles. Compared to estimates from the Uchuu Simulations \citep{2021MNRAS.506.4210I}, for example, of a lower limit of $\sim$1000-3000 particles to calculate halo properties like concentration, our maximum value is theoretically more than sufficient to produce a well-defined potential well for boundedness calculations. Using $\sim$124.7 times fewer particles, in this example, results in 15,556 times fewer calculations when those calculations scale with $n^2$, where $n$ is the number of particles used to solve a halo. This is useful for solving for the potential energy of each particle, and allows us to pursue more robust energy-solving checks on our halos. In our testing, our particle samples are faster to collect and less memory-intensive than building a hierarchy for a \citet{1986Natur.324..446B} algorithm such as used in {\sc Rockstar}  or the TreePM scheme \citep{1995ApJS...98..355X} in GADGET-2 \citep{2005MNRAS.364.1105S}, which theoretically scales as $O$($n_o$log($n_o$)). Using our method for the largest halos, loading the particle data into memory (scales as less than $O$($n_o$), see Sec. \ref{sec: memory}), particle sampling (scales as $O$($n_o$)), and energy-solving (does not scale with $n_o$ since $n$ is capped) take around the same amount of time and together scale less than linearly with the number of particles in a halo (see Sec. \ref{sec: timing} and Fig. \ref{fig:scaling} for solving timings). %These components take a similar amount of time due to the techniques described here and in the following sections that we have employed to address bottlenecks in each of these routines as either of them became the limiting step.

\subsubsection{Energy Calculations}
\label{sec: energy}

We calculate the specific potential and kinetic energies of each particle based on the center of mass and mass-weighted mean velocity of the search region. A calculation of halo boundedness requires accurate foreknowledge of the center of mass velocity of the halo so these initial energies cannot be used to solve for halo boundaries, and we will ultimately need to iterate solutions for the true halo center. However, even without a center to build from, it is possible to identify groupings of combinations of specific kinetic and potential energy with respect to the overdensity center that is spatially co-located and use the centers of those groupings as initial guesses for multiple halos and subhalos in an overdense region.  

To associate particles into these halo-like groupings, we use k-means clustering. These clusters are created by linking data in multiple dimensions, where each dimension represents a different attribute of the data. We use five dimensions to cluster particles: relative particle positions (three dimensions), log potential energies (one dimension), and the ratio between the log kinetic energy to the log potential energy (one dimension). Each of these quantities was normalized by their mean values and centered about zero so that they have roughly equal weight. Fig. \ref{fig:method 1} (top left) shows the energy distribution of a sample grouping of particles found with k-means clustering inside a single overdense region of sampled particles. Even though these halos are spatially co-located, the potential and kinetic energies of their particles form distinct clusters, which are colored by halo. Since the number of dark matter particles can differ by orders of magnitude between halos and their satellites or subhalos, k-means clustering would typically struggle to find smaller halos in a complex due to position-space skewing (the mean tending towards the properties of the more numerous particles of the larger halo). Our particle sampling procedure has the added effect of making it much easier to identify subhalos and merging components by spreading the particle positions much more evenly in space and suppressing the particle density of cores.

Once clusters are identified, for the second iteration, the specific kinetic and potential energy of the particles within 1.2 times the cluster's radius (from cluster center to the distance of the furthest particle within the cluster) are calculated based on the cluster's center of mass and cluster's mass-weighted mean velocity. This step uses a random sample of 5000 of the cluster's particles with up-scaled masses to find bound particles (Kinetic Energy + Potential Energy $<$ 0), from which we define an initial center and radius out to $1.2r_{150c}$, where r$_x$ is defined to be the radius wherein the density of the volume, $\Delta_{c,x}$, is $x$ times $\rho_c$. As a final iteration, using particles within $1.05r_{150c}$, energies are recalculated using the cluster mass-weighted mean velocity of particles and the halo center. Here, we use the energy-weighted center of the bound particles ($\frac{\sum E*\vec{r}}{\sum E}$, where $E$ is the total energy of a particle at position $\vec r$) which we refer to as the ``center of energy'', rather than the center of mass. Where center of mass depends on the membership of particles that are included in the calculation, the center of energy is heavily biased towards the location of the bottom of a potential well, which was found to be more stable over timesteps. Since we calculate particle potential and kinetic energies to calculate boundness, the center of energy is also a much more consistent frame of reference than the location of a density peak, which is systematically offset from both the center of mass and the bottom of the potential well of a halo during a major merger and could heavily skew the calculation of energy.

From within $1.05r_{150c}$ from the center of energy, a final list of bound particles from the full sampled list is determined, then a set of radii are identified from among the bound particles forming a range of overdensites and the IDs of the bound particles within a target overdensity, $\Delta_{c,x}$, are recorded, which ranges from $x=150$ to $x=1000$, when possible. In Fig. \ref{fig:method 1} (left bottom), the halos were found to inhabit a single overdense region at the end of the last iteration. Four halos are identified including one main halo and three subhalos or merging halos that reside within the radius of the main halo. Typically, a density of $\Delta_{c,x} = 200\rho_c$ is used for halos radii, which is an approximation based on linear collapse theory. As such, we enforce a density of at least $\Delta_c = 200\rho_c$ for our initial particle cluster-finding after our overdensity searches. However, we still accept halos with higher and lower overdensities in subsequent steps of the pipeline so long as halos contain more particles than a minimum mass and minimum particle count threshold. We set this threshold to half the minimum dark matter particle mass (effectively zero for this work). For the results of this work, which are based on $\Delta_c = 200\rho_c$, we search for bound particles out to max(1.05$r_{150c}$,1.2$r_{\rm last}$), where $r_{\rm last}$ is the radius last confirmed for the halo in a prior timestep. This allows us to reliably recover $r_{200c}$. In current versions of the code that additionally solve for virial radius, the final bound particle search extends to the minimum of max(1.05$r_{vir}$,1.2$r_{\rm last}$) and 1.05 times the distance from the center of energy to the farthest bound particle, $r_{\rm max}$. The latter half of the minimum, 1.05$r_{\rm max}$, aids the retention of smaller sub-halos that might otherwise be lost in the context of nearby potential wells. When we solve for virial radii, the reported radii are for overdensities that range from $x=80$ to $x=1000$, when possible. This is in addition to the virial overdensity, which can be slightly less than 100 at $z=0$ and may be as high as $18\pi^2$ at high redshift, depending on the cosmology. %For main halos $r_{\rm vir} \approx r_{200}$ and deviations from this are most likely to occur in subhalos or halos that are nearer to the dark matter mass resolution.

Each newly confirmed halo found with this procedure is then reconfirmed by repeating the cluster-finding process, including a re-sampling of particles centered on candidate halos rather than our initial search volumes to properly characterize halos that exceeded the initial search volume as well as to ensure that the definition of the halos was robust against a different initial condition. To reduce redundant halos, clusters are investigated in order of decreasing mass, and particles bound within a much larger halo's $r_{800}$ are excluded from consideration for further clusters.  This allows particles to be bound to multiple halos but only in the extremities of a larger halo. Halos and sub-halos that are within $r_{700}$ of a more massive halo are tracked with our particle tracking algorithm at a different point in our calculation (see Sec. \ref{sec: Back-model}). In this confirmation step, the particles recorded as part of the halo are seeded as one of the clusters whether or not they are part of the particle sample for the energy calculation. A final particle list is recorded from the bounded particles from the combined sample. Because each candidate halo is sampled individually in the confirmation step, small halos and large halos are effectively comprised of a similar number of particles and are similarly well-defined.

Because we set our overdensity threshold for the overdensity-finder to $\Delta_c = 300\rho_c$ and our cluster-finder to $\Delta_c = 200\rho_c$, the same halo-complexes are subject to multiple searches about component halos, resulting in a more complete halo list. For the initial round of cluster-finding after an over-density finding step, the process is repeated for each sub-halo that has a mass over 1/5th the mass of the main halo, which can result in more subhalos and subhalos of subhalos. In the current version of the code, all our target overdensities are replaced by the formalism by \citet{1998ApJ...495...80B}(Eq. 6) for a redshift dependent overdensity $\Delta_{\rm vir}$, including during the overdensity-finding steps, but the results analyzed here are based on $200\rho_c$ or $300\rho_c$ as and where indicated. The effect of lowering this threshold to the virial overdensity is that the initial search produces more low-mass halos. However, because the lowest mass halos are not well-defined in this spherical overdensity context, this change is not scientifically relevant to this investigation and will be explored in future work. Additionally, in this work we set the minimum number of particles in a halo to 11 and reject halos with fewer particles, but in the current version we only require that a halo contains at least one dark matter particle.

\subsubsection{Pruning}
\label{sec: pruning 1}

Results are then pruned to exclude overlapping or redundant halos in a manner similar to the pruning of the overdense regions in the prior steps. A new halo is removed if it meets the following conditions for any pairwise comparison with a more massive halo:

\begin{enumerate}
    \item \textit{Similar mass.} The halo center is within 0.2$r_{\rm vir}$ of the center of the more massive halo or the center of the more massive halo is within the smaller halo's 0.2$r_{\rm vir}$. 
    \item \textit{Similar radius.} The mass of the halo is within 0.2 of the mass of the more massive halo. 
    \item \textit{Similar velocity.} The absolute value of the dot product of the unit-normalized bulk velocity vectors of the two halos minus one is less than 0.05.
    \item \textit{Less consistent.} The halo has been tracked for the same or fewer timesteps than the more massive halo (only applicable when applying this to particle tracking, see Sec. \ref{sec: tracking}).
\end{enumerate}

The third condition is especially important because it discriminates between mergers and alternate or duplicate definitions of the same halo. Due to our energy-solving step, we have added the advantage of knowing the velocity of the particles that are gravitationally bound to each halo, which can be more helpful for halo discrimination than just having the velocity of particles within a spherical overdensity. 

Even after pruning, our cluster-finding procedure typically produces a large initial library of halos. However, not all subhalos and halos that are very near to other halos are included in the initial list, owning to the inherent difficulty of defining hierarchical structures of bound particles.

Because our pruning conditions are limited to comparisons with more massive halos and we can skip comparisons with halos that have already been pruned, the procedure scales better than $O(n_{\rm halo}^2)$, where $n_{\rm halo}$ is the number of halos at a particular timestep, but it can still become time-consuming for large values of $n_{\rm halo}$. Therefore, the cluster-finding results for each timestep are shared between cores, and this step is parallelized.

\subsection{Particle Tracking}
\label{sec: tracking}

Our algorithm includes modes to track halos either forward or backward in time using particle tracking in conjunction with our particle cluster-finding algorithm. Though we can find many subhalos in individual timesteps, we find most of these halos using the combined overdensity-finding and energy-finding process run on different timesteps to catch halos when they are more isolated. Between these timesteps, our particle tracking algorithm maintains the halo catalog by performing the following series of backward and forward modeling procedures, which work in conjunction with energy-solving (Sec. \ref{sec: energy}).

\subsubsection{Backward-Modeling}
\label{sec: Back-model}

Backward particle tracking begins with projecting the halo center of mass quadratically using the halo velocity and its numerically calculated acceleration ($\Delta v/\Delta t$). From this center, we first test a radius of either 3.5$r_{\rm last}$ or three times the velocity of the halo multiplied by the time difference between timesteps, whichever is larger. This sphere that ensures we capture most, if not all, the known particle IDs of the halo in the new timestep. Using the center of mass of the known particle IDs, we further cull the particles to 1.75$r_{\rm last}$ about the center of mass of the matched particle IDs. Culling the particles in this way brings our sample region of interest back to the scale of the halo. We include both the sampled particle list and up to 5000 randomly selected known particle IDs from the source halo to create a final particle list for our halo search. Then, our iterative cluster-finding is performed to find a main-antecedent halo and any new progenitors or sub-halos. 

Our algorithm often finds more than one candidate progenitor for each halo, so we use a cost function to determine which halo is likely to be the true progenitor.

Our cost function is:
\begin{equation}
    {\rm Cost} = |r_{\rm mass}-1| + c_1 / f_{\rm IDs}^2 + c_2 |cos(\theta)-1|+ c_3r_{\rm dist},
\label{eq:cost}
\end{equation}

where $r_{\rm mass}$ is the ratio of the source halo mass to the candidate progenitor mass, $f_{\rm IDs}$ is the fraction of particles in the source halo that are in the candidate progenitor, $\theta$ is the angle between the center of mass velocities of the bound particles of the candidate progenitor and the source halo, and $r_{\rm dist}$ is the distance between the projected position of a halo from the antecedent timestep using its acceleration and velocity and the center of the candidate progenitor all divided by the antecedent halo radius. Thus, all components of Eq. \ref{eq:cost} are ratios or normalized to be independent of units. For our results, constants $c_1$, $c_2$, and $c_3$ are set to 10, 1, and 100 respectively. This fiducial order of magnitude was selected by observing the smallest change between timesteps in any of these parameters as halo progenitors were selected. Then, those values were used to chose the appropriate order of magnitude of $c_1$, $c_2$, and $c_3$ such that each factor is roughly equally weighted. Therefore, the solution is affected if any of these factors deviates from this minimum. After testing each candidate, we only accept the candidate with the lowest cost as long as it meets a minimum threshold, ${\rm lim}_{\rm IDs}$, as well as a value of $|r_{\rm mass}-1|$ less than a maximum threshold, ${\rm lim}_{\rm mass-diff}$ as given in Tab. \ref{tab:limits}. 

Initially, the density confirmed in the source halo, $\Delta_{c,x}$, is used as a target for the final overdensity in its progenitor. This allows us to be consistent in our tracking of halos that are only well-defined at high overdensities (ex: merging halos or subhalos) or low overdensities (ex: tidally disrupted halos).

If an acceptable progenitor is found, the other candidate halos are retested with the cluster-finding algorithm. If they are confirmed, they are accepted as either co-progenitors if they share particles with the source descendant halo or new halos if they do not. Sets of new halos, antecedent halos, and co-progenitors are accepted only if a main antecedent halo is confirmed using cluster-finding. 

If an acceptable progenitor is not found among the candidates, $\Delta_{c,x}$, ${\rm lim}_{\rm IDs}$, and ${\rm lim}_{\rm mass-diff}$ are changed for a second try  or third try with values shown in Tab. \ref{tab:limits}. The values in this table are for the most recent version of the algorithm. Each round is initiated in an order that prioritizes a class of results. Round 1 is a halo with a consistent mass and composition with a overdensity that is consistent with theory \citep{1998ApJ...495...80B}. Round 2 is a halo with a overdensity that is consistent with the prior antecedent halo properties with an allowance for a larger change in halo mass. Round 3 is targeted towards rarer situations such as major mergers where the composition and mass of the halo changes drastically between timesteps. Irrespective of the limits and target overdensity, the halo with the best cost function within the limits is selected as a progenitor.

The values in Tab. \ref{tab:limits} are under active development to try to cover the widest range of edges cases as we continue to test our procedure on more simulations. In terms of robustly tracking smaller halos, the values presented in the table represent an improvement from those used to produce the results analyzed in this work, which act as a baseline for further adjustments. In the current version of {\sc Haskap Pie}, the target overdensity in the first round is set so that it is biased towards the target overdensity and in the analysis presented in this work that bias is towards $200\rho_c$, which replaces $\Delta_{vir}$ in the formula. Additionally, results reported in this work were determined using two rounds (first and third round) instead on three used in the current version. Analyses based on these and other improvements to the algorithm will be documented separately.

\begin{table}[]
    \centering
    \begin{tabular}{c|c|c|c}
        Round & Target Overdensity & ${\rm lim}_{\rm IDs}$ & ${\rm lim}_{\rm mass-diff}$ \\
        \hline
         1 & $\Delta_{c,x}-$5$\times$sign($\Delta_{c,x}$-$\Delta_{vir}$) & >0.85 & <0.15 \\
         2 & $\Delta_{c,x}$ & >0.85 & <0.35 \\
         3 & $\Delta_{c,x}$ & >0.55 & <0.5 \\
         \hline
    \end{tabular}
    \caption{Values used in conjunction with Eq. \ref{eq:cost} as targets and limits for matching halos between timesteps. Limits are initially tight as the general preference is for consistent halos, but limits are gradually relaxed to cover more edge cases.}
    \label{tab:limits}
\end{table}

If cluster-finding fails to find a suitable candidate after all rounds and the halo had been previously tracked with cluster-finding for three consecutive timesteps, we employ a different method for particle tracking. Using the known particle IDs, we assume a spherical potential based on the center of mass of the tracked particles from the descendant halo. This differs from the cluster-finding algorithm in that we can very quickly determine particles bound to a spherical potential since it does not depend on an iterative free search for bound particles. This method is much more likely to find bound particles, however, using this algorithm creates less robustly defined halos. If this method still fails, our last resort is to construct an overdense region to about the center of mass of the particle IDs.

In subsequent timesteps, we always start with cluster-finding before again resorting to these methods and we place limits on how many times spherical energy-solving or overdensity-solving methods are used in cases where cluster-finding and energy-solving solutions should be calculable before a halo is declared lost. If a halo is inside a much larger halo, we allow this algorithm to run consecutively but track the number of times it is used. However, if no nearby halos are more massive, we limit the times we use this form of energy solving in consecutive timesteps to a total of $\sim$1/10th the timesteps in backward modeling or forward modeling. About 10\% of timesteps along halo tracks are solved using these methods in our N-body (dark matter only) simulations and around 15\% in the cosmological hydrodynamic zoom-ins with many sub-halos.   Even if progenitor halos are found, they are subject to the pruning procedure described in Sec. \ref{sec: pruning 1} at every time step. In any version of particle-tracking, we require that halos have more than five particles for this work and at least one particle in the current version.

%To accommodate situations where halos may undergo mergers, close passes, or other forms of disruption before settling, the particle IDs of the target halo are only compared to the id list from several antecedent timesteps before or after the present, depending on whether the algorithm is forward-modeling or backward-modeling the particle tracking.  If a halo meets these conditions but is smaller than a preset threshold for halo mass changes and our particle tracking is going backward in time, it is recorded as a possible co-progenitor. Halos of any mass that are not associated with any antecedent halos are recorded as new halos. 

% If one is not confirmed, the particle algorithm adjusts the search volume radius, rejection mass tolerance, and number of k-means clusters for several iterations until one can be confirmed, before declaring the search a failure. This algorithm is tuned conservatively to ensure that an antecedent halo can be identified and usually returns redundant co-progenitors, which are pruned.  

Backward modeling is initially used to populate the halo list with co-progenitors and new halos. With our method, halos are usually identified when they are less co-incident with other halos, but it is our preference to track halos down through their mergers as they are being tidally disrupted and for halo lists to extend for as many timesteps as a halo radius can be defined. 

%While it is our preference that all halos are confirmed in antecedent time steps as a bound system, we also allow halos that have been successfully tracked between timesteps a budget of four instances of being tracked from the particle list alone as long as a virial radius can be calculated. 

\subsubsection{Forward-Modeling and Tree Pruning}
\label{sec: final Pruning}

\begin{figure*}[t!]
\begin{center}
\includegraphics[width=\linewidth]{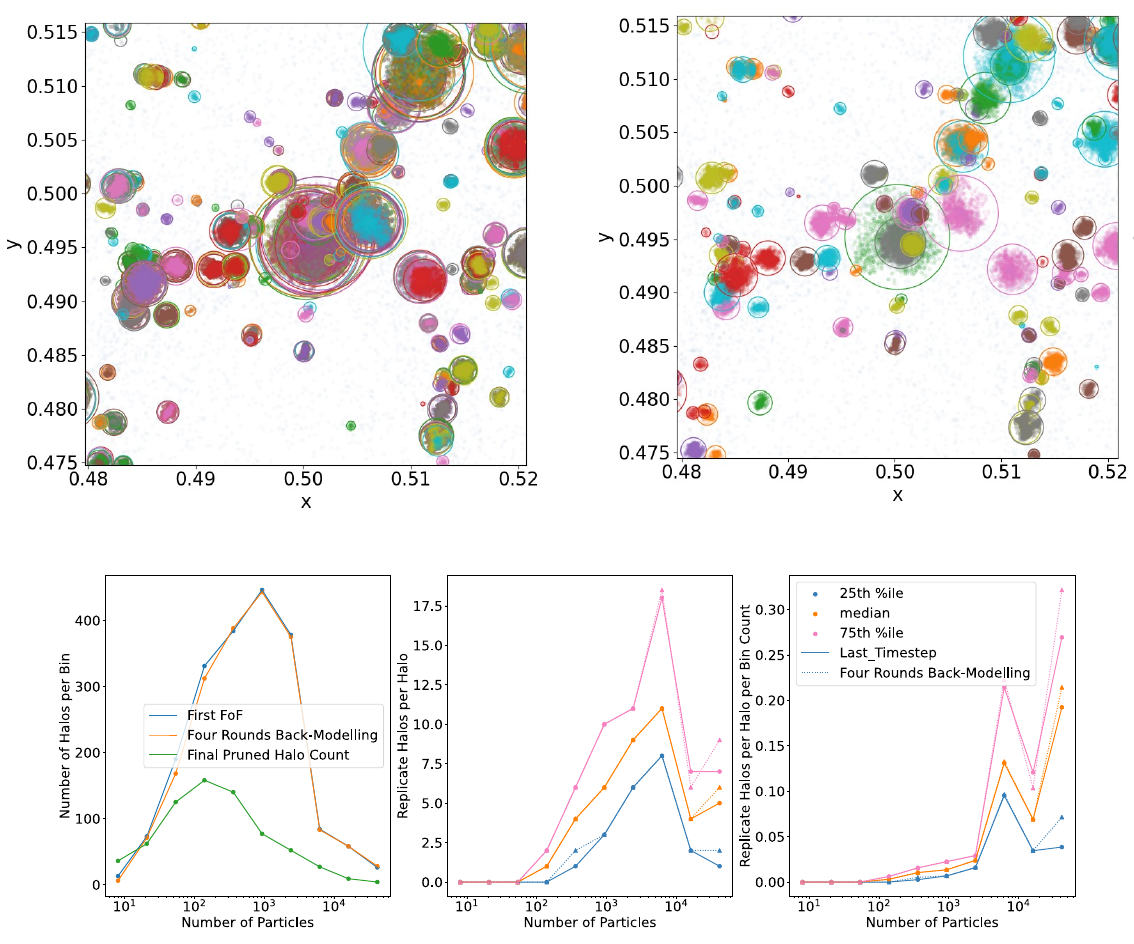}
\caption{Plots demonstrating halo redundancy for our test simulation. Top left: All halos found by our cluster-finding solution within a four virial radii box centered on the main halo ($1.3 \times 10^9$ M$\odot$ at $z\sim7.5$) before pruning, showing multiple solutions for most halos and sub-halos. Colors indicate the radius (200c) and sampled particle membership of corresponding halos. Top right: The halo catalog after four rounds of tracking and pruning. Bottom row: Left: The total halo counts from the combination of our overdensity-finding and energy-solving method, without pruning (blue), after four rounds of particle tracking without pruning (orange), and the final halo counts with prunning throughout the calculation (green) all versus halo particle counts. Center: The inter-quartile range of number duplicates that are produced for each halo before (solid line) and after (dotted line) particle tracking without pruning versus halo particle counts. A value of zero indicates halos are singular and not duplicated. Right: The same as the center plot but divided by bin counts to normalize duplicates by halo populations. For halos with at least 100 particles, the number density of duplicates increases with particle count with an average of more than seven duplicates per halo. The number of duplicates is therefore not limited by the particle-tracking algorithms.}
\label{fig:duplicated}
\end{center}
\end{figure*}

To complete the halo tree, after several timesteps of backward-modeling, the overdensity/halo-finding step is repeated to further populate the halo list, two more backward-modeling steps are performed to confirm new halos, and then we forward-model for several timesteps while pruning for redundancy. During forward-modeling, we run the particle cluster finding routines and follow the backward-modeling procedure, except that we only confirm halos in earlier timesteps that are not present in later timesteps and do not confirm or add new halos to the list. Halos that cannot be tracked for five timesteps are completely removed from the halo list and future rounds of forward and backward modeling are used to find and track better candidate halos. This five-timestep threshold for removal was found to prune the tree in such a way as to give precedence to the longest, most well-defined portions of a halo track and allow the algorithm to build the halo tree outward from there. 

A final round of forward-modeling is performed on each timestep sequentially from the earliest to latest to extend all shorter-lived halos forward in time, if possible. When all timesteps have been forward- and backward-modeled, the algorithm completes by rebuilding the halo tree. First, all short halo histories are again removed. The number of halos used to define a short history is either five or the total number of snapshots minus two, whichever is less. Halos that begin and end within $0.8r_{\rm vir}$ of a halo with twice as long or greater of a track are also removed. Then, halos that have their latest timestep inside another halo ($r < 0.8r_{\rm vir}$) with a higher mass ($M < 0.8 M_{\rm larger}$) and sharing any tracked bound particles are assumed to have merged at the halo's final timestep. 

Halo names are then selected to connect each halo to their descendants and progenitors. The largest non-merging halo when comparing each halo's final mass is named halo `0' and each smaller non-merging halo is numbered `1'-`N-1' in order final mass. Halos that merge are named after the halo they merged into as well as in order of their merger from the last timestep. For example, the last halo to merge into halo `0' is halo `0\textunderscore 0', the second is halo `0\textunderscore 1' and the third halo to merge into halo `0\textunderscore 1' is named `0\textunderscore 1\textunderscore 2' and so on. If multiple halos merge in the same timestep, the larger halos are counted first. The main progenitor branch along the tree is therefore usually, but not always, based on the highest-mass progenitor. That connection is common in tree-making algorithms, but in our method, it is sensitive to the conditions described in our progenitor tracking routine and cost function. This will be explored in future work while we refine our definition of mergers and study them using our algorithm.

\subsection{Parameter Selection and Redundancy}
\label{sec: parameters}

Our algorithm hosts a number of modeling and parameter choices that were selected to serve the dual objectives of creating a complete sample of halos as well as respecting computational limitations in processing power and memory availability. To ensure that our cluster-finding choices did not impact the demographics of our resulting halo catalog, we built in a high level of redundancy. This redundancy comes in the form of duplicate realizations of halos that serve as candidates for our final reported halo. Here we define ``duplicates" as halos that fail the first three pruning conditions enumerated in Sec. \ref{sec: pruning 1} with the caveat that the comparison is not only to the more massive halo. Duplicates are systematically produced from overlapping k-means clustering searches in a halo-containing volume every time we run our energy-solving procedure (see Fig. \ref{fig:method 1} for an isolated example).

If there are a sufficiently large number of duplicates for the sample of halos, the algorithm is not minutely sensitive to parameter changes since our final results are abstracted from those changes through redundancy. Therefore, we focus our testing on confirming the presence of duplicates while we make parameter and algorithmic choices. Therein we seek to maintain the robustness of our results as we simultaneously optimize over memory usage as well as CPU time. Because there are many interconnected elements of our algorithm, it is nearly impossible to isolate the effect of each change or choice so we followed a decision-making procedure that ensured the algorithm would always tend towards a more robust solution as we developed {\sc Haskap Pie}. That process involves running our algorithm on our test simulation for each change in configuration or tuning and confirming that redundancy is maintained. If a solution is no longer as redundant and robust, that change is reversed and further changes to that element are frozen until the algorithm evolves sufficiently to warrant a retest. In several cases, changes to the algorithm overlaid new redundant procedures on existing ones. New procedures were retained indefinitely unless they caused a significant increase in resource utilization or if they came into conflict with another element of the algorithm in such a way as to reduce robustness. This overall process was repeated over 1,400 times during the development of {\sc Haskap Pie}. Thus our development strategy was not a parameter optimization, but a process where new methods and ideas were continually introduced and tested. Due to this process, the algorithm became more robust over time as all accepted changes either increased redundancy or kept it constant.  This resulted in a complicated, but comprehensive algorithm that simultaneously employs multiple halo finding and tracking techniques. These techniques compete through the cost function (Eq. \ref{eq:cost}) and through our pruning conditions. Several of the sub-algorithms that passed testing, such as the particle sampling procedure, have the dual benefit of providing optimization and improving our ability to locate and track halos. Associated decisions, such as the number of annular sectors, were made incrementally after confirming that each choice increased or did not reduce the robustness of the solution.

In Fig. \ref{fig:duplicated}, we show the distribution of duplicates that arise from the scientifically mature version of {\sc Haskap Pie} to demonstrate the performance of the algorithm with respect to this metric. In our test simulation, as shown in Fig. \ref{fig:duplicated} (top left visually and bottom row statistically), there were over seven duplicates for each halo on average with the number of duplicates rising with both particles count and the number density of halos of a similar size in the simulation volume. For halos with at least 100-200 particles, multiple realizations are present for our algorithm to discriminate between and so our final halo counts similarly peaks between 100 and 200 particles (Fig. \ref{fig:duplicated}, bottom left, green line).

As further discussed in Sec. \ref{sec: results}, halo demographics were not complete for halos with smaller particle counts. Below 100 particles, halos are increasingly difficult to define, locate and track. Thus, several of the tracking procedures described in Sec. \ref{sec: tracking} are aimed at improving the survivability of low particle number halos with irregular shapes and poorly-defined gravitational potentials, with diminishing returns. As a result of those procedures, the final halo number distribution for low ($<$20) particle halos is slightly higher in our final results than in the initial search.

For larger halos, many of these choices and procedures are completely redundant and have no effect on the survivability of halos. As shown in Fig. \ref{fig:duplicated}(orange line in the bottom left plot and dashed lines in the center bottom and right bottom plots), we are able to track each duplicate individually and tracking in general does not limit our redundancy. Conversely, due to the use of energy-solving and k-means clustering in every time step, the number of duplicates increases as the algorithm progresses without any additional overdensity finding. This is especially the case for halos with at least 10$^{4}$ particles. This also supports our choice to restrict particle sampling to regions with more than 10$^{4}$ particles as we can expect good redundancy and therefore robust solutions in that range.

Because there are several realizations of each halo available to build our catalog and merger tree, the most influential choices we made are in the \textit{pruning algorithm}, which we have intentionally kept simple, neutral, and non-controversial. The current pruning conditions are, in effect, only defining the uniqueness of a halo.  As shown in Fig. \ref{fig:ART Shinbad} (top right versus top left), our pruning is sufficient to collapse the solution into a singular distribution of halos without removing viable, well-defined halos from the catalog. As is evident from plots shown in Sec. \ref{sec: results}, our pruned results are physically singular, and each halo and sub-halo in a complex system displays a unique orbit, mass, and relative angular momentum. Separately from our redundancy testing, we test the completeness of the resulting final halo distribution by comparing our halo mass function to theory and other halo-finding results as presented in Sec. \ref{sec: n-body}(Fig. \ref{fig:Nbody HMF}) and Sec. \ref{sec: AGORA}(Fig. \ref{fig:AGORA HMF}).

\subsection{Algorithm Optimizations and Run Timings}
\label{sec: timing}

\begin{figure}[h]
\begin{center}
\includegraphics[width=\linewidth]{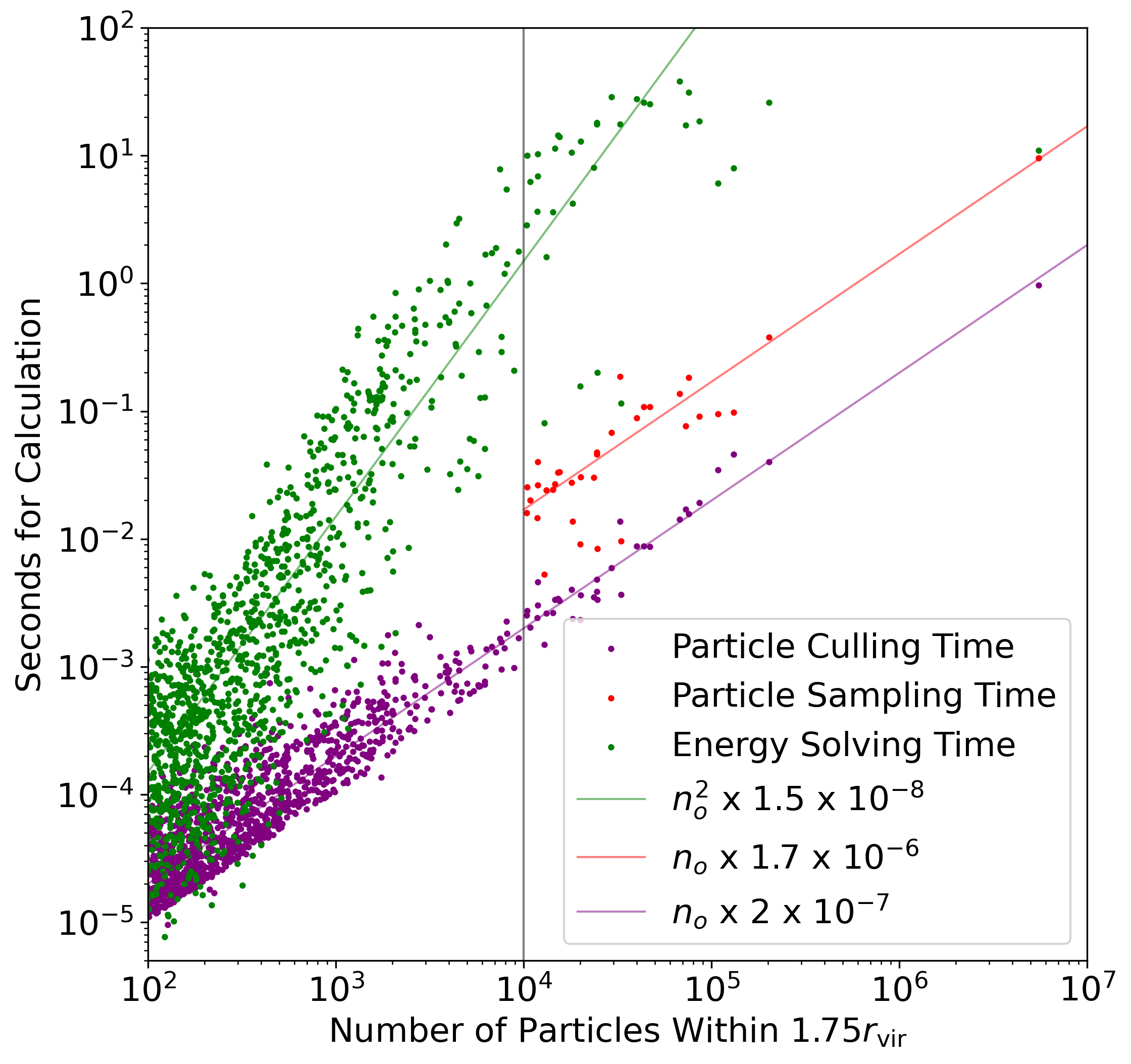}
\caption{Example per-halo single-core computational time for a sample of halos in a timestep showing the three most expensive portions of the algorithm that threads must calculate independently: particle pruning (purple), particle sampling (red), and energy solving (green). Lines are also drawn showing hypothetical linear or quadratic relationships for each of the calculation types of the corresponding color. Note that particles are only sampled when the number of particles in the search region is greater than 10$^4$ (shown as a vertical black line). The median and mean total calculation time (all three combined) per halo in this sample is $\sim$0.7ms and $\sim$0.4s respectively. Particle sampling is projected to become the slowest calculation for halos with more than a few million dark matter particles unless further optimized.}
\label{fig:scaling}
\end{center}
\end{figure}

Due to the popularity, accessibility, and user-friendliness of Python, and \texttt{yt}'s Python-based package of simulation and data visualization tools, we chose to develop our algorithm entirely in Python and aim to integrate the peer-reviewed version of our routines into \texttt{yt} to promote broader adoption, transparency, and open development. However, Python is not as efficient as languages that reside closer to machine code and most other halo-finding codes run natively in faster coding languages like C and FORTRAN. That includes RCT, which is connected to \texttt{yt} through a separately maintained front-end wrapper. Despite being Python-based, we optimized our code with a target of making it capable of solving science-scale halo trees on a typical laptop in a reasonable amount of time.

Figure \ref{fig:scaling} shows routine run timings for a sample of halos in AGORA's GADGET-3 simulation at $z=0.3217$, broken down into the three most expensive single-threaded calculations: energy-solving, particle sampling, and particle culling. Additionally, the time to initially loading particle data into memory can be significant. For the purposes of timing comparisons, we define the entire k-means clustering and boundness-finding routine described in Sec. \ref{sec: energy} including multiple iterations for the main halo and any subhalos caught with the clustering routine, but excluding the confirmation steps that re-sample the particles to fully solve nearby halos as a single instance of ``energy solving". We find that energy-solving is the slowest step below a few million particles and particle sampling is projected to become the limiting step at higher particle counts. Energy calculations are initially $O(n_o^2)$ and then hit successive caps on particle counts due to sampling in the energy-solving iterations as well the whole particle list that limit and even partially reverse scaling. At five million particles, our procedure saved five orders of magnitude in computational time for energy solving as compared to the plotted $O(n_o^2)$ scaling relationship.  %and two orders of magnitude as compared to a hypothetical $O(n_o$log$_{10}(n_o$)) scaling relationship from $n\geq3000$. 
Culling times scatter upward at low particle counts because smaller halos may be pulled from denser regions that require more calculations. Per-halo particle loading timings are more complex to calculate because of our memory optimization strategy (see Sec. \ref{sec: memory}), but can be a bottleneck in our pipeline.

The largest simulation we have run our halo finder on was an ENZO 512$^3$ N-body simulation described in Sec. \ref{sec: sims}. On the Texas Advanced Supercomputing Center's Stampede3 Skylake-based nodes, for a typical back-modeled timestep in that simulation, 11,000 halos were solved in 303 seconds on 80 cores (24,240 core-seconds, 0.027s/halo), but took an additional 771 seconds (0.097s/halo/timestep total) to load region data from \texttt{yt}. For the AGORA's ART-I zoom-in simulation, 1,100-halo timesteps were solved in about 77 seconds on 80 cores ($\sim$ 6,160 core-seconds) of which 15 seconds were used to load regions (0.07s/halo/timestep total). Both simulations have similar resolutions and maximum halo sizes of order $10^{12}$ M$_\odot$ (Milky Way-sized) containing several million particles and several dozen subhalos. Forward-modeling is typically slower per halo as energy solutions are more likely to fail for halos that were not previously detected in back-modeled steps and most forward-modeled halos are within the particle-dense regions within and around the largest halos. In ART-I, a typical forward-modeled timestep for the same simulation completes at a rate of 0.14s/halo/timestep whereas the ENZO N-body simulation achieves about 0.27s/halo/timestep. The ENZO high-$z$, high resolution zoom-in simulations used in \citet{2024ApJ...969..144S} ran forward and backward-modeling at a rate of about 0.07s/halo/timestep on a ten-core laptop including about six seconds per timestep to load regions. This is notable because this simulation contains 40 times fewer particles than the AGORA simulations in the largest halos. When run with the same computational configuration, AGORA takes less than twice as long per halo per timestep.

\subsubsection{Memory Optimizations and Load-Balancing}
\label{sec: memory}

Per-core memory usage and inter-core communication times were found to be a significant bottleneck on computing clusters whereas smaller core counts were found to limit performance on laptops. The algorithms are written such that simulation data, including full particle information (positions, masses, velocities), is never shared between cores and the sampled particle IDs are only shared as needed. This is achieved by organizing the job scheduling prior to particle-tracking each timestep so the relevant simulation data is only loaded once, from which metadata is extracted and excess data is quickly removed from memory. This is further optimized by using the same particle-loading step to load all subhalos regions wholly within the region about a larger halo in the same step without requiring any additional memory reading. 

To perform load-balancing during the particle cluster-finding process, we identify the projected search volume for halos as described in Sec. \ref{sec: Back-model}. Then, in order from the largest to smallest volumes, we check how many overlapping search volumes are entirely contained in each volume. If more than min(10,$n_{\rm procs}$) halos are inside a volume, where $n_{\rm procs}$ is the number of available processors, we split the volume into 3$^3$ sub-volumes and identify the sub-volumes with interior search volumes. We use 3$^3$ instead of an octree splitting so that one of the sub-volumes is centered on the center of the halo, where there are likely to be subhalos. This process is repeated recursively for each new sub-volume with more than min(10,$n_{\rm procs}$) interior search volumes except the sub-volumes are split into 2$^3$ sub-sub-volumes after the first splitting. This effectively builds a hierarchy of volumes that each have a limited number of interior search volumes.

Then, the combined list of volumes and subvolumes are partitioned into groupings of equal cumulative volume based on the number of halos and the number of cores to ensure that each core is allocated a roughly equivalent volume of the simulation as well as a roughly equivalent number of halos to solve. Depending on the number of available cores, particle data and halo-solving for these groupings of volumes are run in sequential rounds so that a limited volume of the simulation is loaded and analyzed simultaneously, saving on memory usage per core. After loading particle data for a group, groupings are further split into sequential batches of five halos per core in such a way that cores are tasked with analyzing their largest halos first, which allows halo-solving to progress quickly after the largest halos are solved. After each batch, halo data is compiled, the cores are code line-synchronized, and necessary data transfers between cores are performed before continuing to the next batch.

In forward modeling, large halos are usually already accounted for and do not need to be loaded so grouping halo searches into common data volumes is less useful. Therefore, if there are more than 1000 halos to forward model, we also check for halo search volumes within the 3$^3$ subvolumes about the largest solved halos and again build a hierarchy from sub-volumes with min(10,$n_{\rm procs}$ and retain subvolumes with at least 5 halos, discarding the rest. This has the effect of reducing loading times and memory usage as compared to individually loading a large number of partially overlapping regions of subhalos.

During a halo-finding round, each core retains a copy of all the particle data from the search regions it has been allocated for the round. When solving a halo search volume within a larger volume, a spatially culled copy of the particle data about the appropriate halo search volume is briefly stored before it is analyzed by our particle summation technique (see Sec. \ref{sec: sampling}), which greatly reduces the number of particles used for cluster-finding and energy solving.

The halo tree is shared and synchronized between cores but is sent in manageable chunks at the end of the timesteps to limit excessive duplicate copies during the transfer process. The large, full list of particle IDs is only stored by a root core which saves halo trees and particle IDs to disk every few timesteps with a backup saved half as often in case of a crash during the writing process. We found that storing copies of the particle ID list on every core would otherwise often become a memory bottleneck.

In order to remove the need to communicate simulation data between cores, all three of the longest operations in our algorithm (particle loading, particle sampling, and cluster-solving) are single-threaded for individual halos and run simultaneously on multiple cores. Further optimizations to the sampling process or a new parallelization strategy may further improve performance. However, our focus on removing memory bottlenecks has also allowed us to use more cores and increase the number of halos we can simultaneously solve so we have balanced these considerations in our current optimization strategy.

These techniques generally reduce the number of particles loaded by a factor of a few, depending on halo positions (2 in the example of AGORA ENZO's $z=0.0898$ timestep in initial backward-modeling, which improved overall loading times by the same factor). As a result, data loading scales slower than the sum of particles in all halos $O(\Sigma n_o)$ for the entire simulation data set. Of note is that most of these memory optimizations are applied to later versions of the code than were used for the run timings calculated for the previous section. Memory-optimized versions of {\sc Haskap Pie} were used on a computing cluster for GEAR, GIZMO, and CHANGA, but our data-handling strategy was different for those data (presaving of particle data to address \texttt{yt} reading discrepancies, see Sec. \ref{sec: sims}) and so we cannot fairly compare their run timings with our other timing data. Ongoing development will lead to more optimizations or functionality over time. 

\section{Results}

\label{sec: results}

\subsection{Halo Mass Function}
\label{sec: n-body}

We compare the number density of halos across mass bins to classical linear theory by roughly following the \citet{1974ApJ...187..425P} formalism of linear theory. Our reported mass function is in terms of: 
\begin{equation}
   dn/dM = 2\frac{\rho_c}{M}\frac{\delta_c}{\sqrt{2\pi}\sigma^2}e^{\frac{\delta_c}{s\sigma^2}}\left|\frac{d\sigma}{dM}\right|,
\end{equation}
where $\delta_c$ is the collapse overdensity of 1.69 given by linear theory, M is the mass of the bin and 
\begin{equation}
    \sigma^2 = \frac{1}{2\pi^2}\int_0^{\infty} P(k,z)W^2(kR)k^2 dk,
\end{equation}

where $P(k,z)$ is the cosmological matter power spectrum evaluated at mode $k$ and at redshift $z$ calculated using CAMB\citep{2000ApJ...538..473L} and the function $W(kR)$ is the Fourier transform of a 3-D top hat window functions for scales of $R$ corresponding to enclosed masses of $M$.

\subsubsection{N-Body}

\begin{figure*}[t!]
\begin{center}
\includegraphics[width=\linewidth]{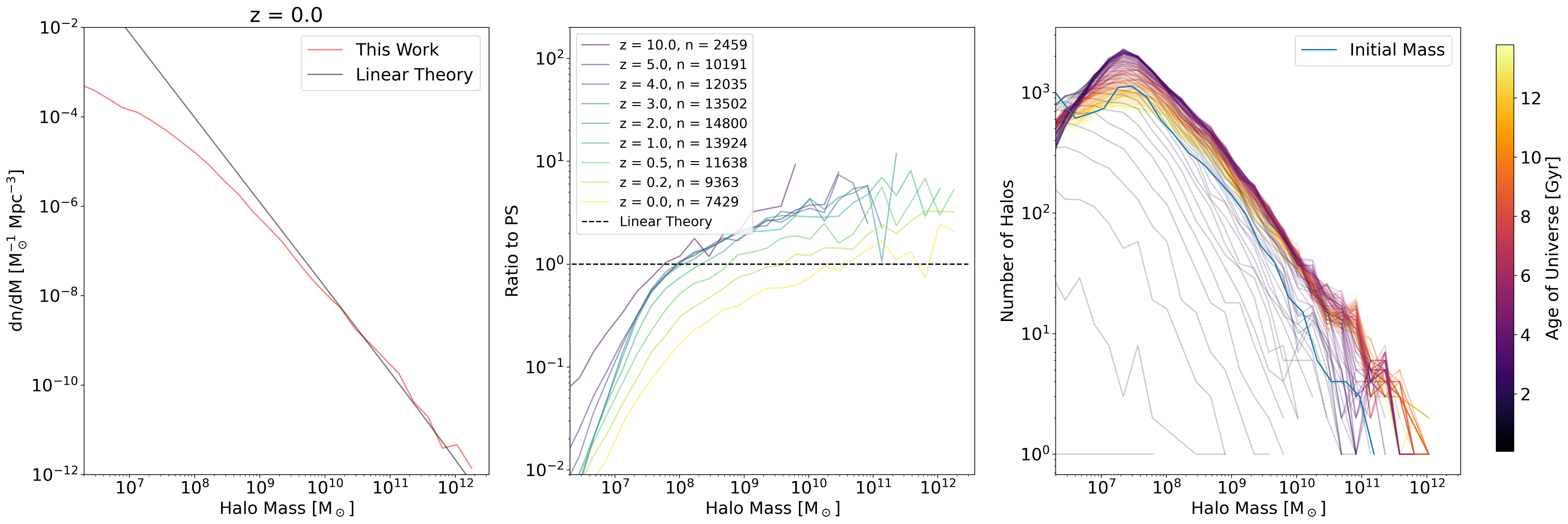}
\caption{Halo populations from a N-body only simulation solved with {\sc Haskap Pie}.  Left: Halo mass function showing linear theory in black and our population in red for simulation data representing $z=0$.  Center,  halo populations across redshifts as a fraction of linear theory \citep{1974ApJ...187..425P} with total halo counts at each included redshift included in the legend. Right: Halo counts as a function of halo mass and is colored by the age of the Universe in a color gradient. The blue line shows the initial mass function of halos. Results show that our results are most complete for halos consisting of more than 100 particles as well as trends associated with halos assembly.}
\label{fig:Nbody HMF}
\end{center}
\end{figure*}

Our algorithm does not presume a force resolution, a linking length, or a softening length and so halos are reported and pruned based solely on whether they are tracked between time-steps as self-bound in addition to the pruning conditions described in Secs. \ref{sec: pruning 1} and \ref{sec: final Pruning}. In practice, a large number of candidate or transient halos are rejected from the initial overdensity-finding and particle tracking calculation, so care needs to be taken to construct a fair comparison to raw halo lists from finders that are not pruned based on halo survivability over timesteps. Therefore, before comparing our results to other halo-finders, we compare our results for the 512$^3$ N-body simulation described in Section \ref{sec: sims} to the \citet{1974ApJ...187..425P} formalism to explore the completeness of an unbiased halo mass function.

 %While this did not have a drastic effect on the number of halos at $z<5$ as these halos were quickly absorbed into main halos, they did have an effect on the mass function at high redshift, which is reflected in the upward-shifted plot of the relative mass function at $z=10$ in Figure \ref{fig:Nbody HMF} (center). 

As shown in the relative halo mass function (mass function divided by the linear theory prediction) in Fig. \ref{fig:Nbody HMF} (center), at intermediately high redshifts ($ 5 \ge z \ge 2$), the results are similar with essentially complete solutions for masses greater than $\sim 10^{8}$ M$_\odot$ ($\sim 1000$ particles). At lower redshifts, two factors contribute to a dip below linear theory as shown in Fig. \ref{fig:Nbody HMF} (left). First, mergers absorb smaller mass halos and shift the distribution lower resulting and are not a part of that halo mass function formalism. However, it is key to note that these values will shift considerably based on the ability of a halo-finder to durably track sub-halo orbits, with longer tracking resulting in fewer mergers and higher numbers of halos at any given redshift. 

The second factor is the free fall time of new halos forming at late times. Taking the usual definitions of the cosmological constants for the determination of $\rho_c$, and the gravitational constant, $G$, the collapse time corresponding to an overdensity of 1.69 is

\begin{equation}
    \label{eq: tff}
    t_{ff} \approx \left(\frac{3\pi}{32 G (1+1.69)\rho_c}\right)^{1/2}.
\end{equation}

At $z=10$, this is $\sim 76$ Myr, but at $z=0.2$, for example, this grows to greater than 2.16 Gyr. Therefore, while volumes of the simulation may reach the critical density to collapse according to the matter power spectrum, and are counted in linear theory, their delayed virialization causes a deficit of small halos that becomes more pronounced as the simulation evolves.

The number of halos above $1 \times 10^{6}$ M$_\odot$ ($\sim 8$ particles) is included with the legend for Fig. \ref{fig:Nbody HMF} (center) and shows that total number of halos grows from high redshift, peaks at $z \sim 1.69$ ($n= 14,939$), and declines as displayed in Fig. \ref{fig:Nbody HMF} (right), which shows a histogram of halos by mass for all timesteps saved in the simulation. 

\begin{figure*}[t]
\begin{center}
\includegraphics[width=\linewidth]{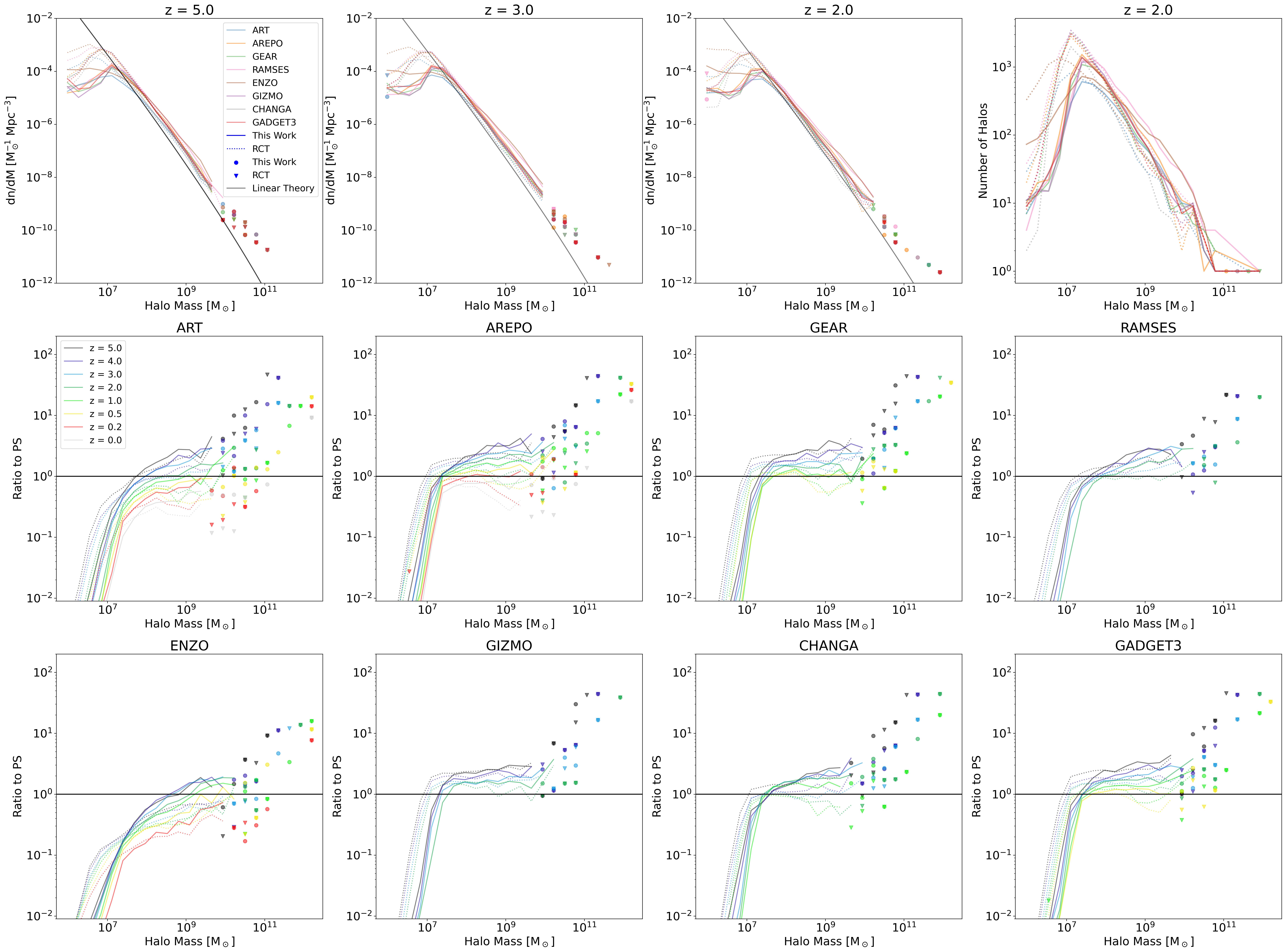}
\caption{Halo mass functions for the AGORA simulations run on with the same initial conditions with eight different codes as labeled and plotted in the same manner and definitions as Fig.  \ref{fig:Nbody HMF}.  Our halo finder consistently exceeds results for RCT when enough particles are present to be captured by our energy-solving and particle-tracking methods.  There are no gross discrepancies in the results for the various codes, which indicates that our method is successfully generalized and performative on the codes' data sets.}
\label{fig:AGORA HMF}
\end{center}
\end{figure*}

Below $\sim 10^{8}$ M$_\odot$, the confluence of the sensitivity of the results to the merger rate, which is a function of the quality of sub-halo tracking and merger dynamics, as well as long free-fall times complicates an analysis of completeness at low redshift. At higher redshift (first billion years), where the merger rate density is lower and collapse times are short, Fig. \ref{fig:Nbody HMF} (right) shows that the number of halos peaks at less than $\sim 10^{7}$ M$_\odot$ ($< 84$ particles), which means that resolution effects are not manifesting as a peak in the number of halos above about 100 particles.

The redshift with the highest number of halos in each mass bin in Fig. \ref{fig:Nbody HMF} (right) is a strong function of halo mass for halo masses $< 10^{9}$ M$_\odot$. This redshift decreases monotonically from $z\sim7$ for $\sim 10^{6}$ M$_\odot$ to $z\sim1$ for $\sim 10^{9}$ M$_\odot$. For higher masses, the curves essentially overlap, and the stochasticity of the matter power spectrum overtakes the trend. This trend can be partially explained as due to halo growth and assembly moving any peak rightward as well as halo destruction (mergers or dissipation) overtaking halo formation at the low mass end as the simulation progresses towards $z=0$. 

In addition to delay time effects, the halo formation rate is potentially limited by the mass of a halo at the time of first inclusion in our halo trees, which may be affected by mass resolution. Halos typically form with a mass less than $10^{7}$ M$_\odot$ (13,421 out of 19,599, 68.5\%) with a median formation mass of $1.3 \times 10^6$ M$_\odot$ (11 particles), which is the minimum we use in our energy-solving step. This means that the histories of a majority of halos are limited by the minimum particle number and that resolution is not limiting the halo tracking itself. However, the blue line in Fig. \ref{fig:Nbody HMF} (right) shows halos can have higher masses at the earliest timestep they are tracked, with a pronounced local maximum at $\sim 5 \times 10^{7}$ M$_\odot$ ($\sim 400$ particles). Though most halos are formed in the first billion years (10,253 out of 19,599, 52.3\%), formation masses grow with redshift and the bump corresponds to the typical formation mass at the redshifts and masses with the highest halo populations in our results ($ 5 \ge z \ge 2$).

\subsubsection{Larger Halos}

While we are primarily focused on Milky Way-sized systems in this work, we also briefly examined the efficacy of our code when studying larger halos, such as those with masses on the order of $10^{15}$ M$\odot$. These halos may contain galaxy clusters rather than main galaxy systems, and act as a test of the flexibility of our technique. We ran {\sc Haskap Pie} on adiabatic simulations used to study cosmic ray accelerations in \citet{2025arXiv250310795S} and we found that halo populations smoothly extended to $\sim1.3\times10^{15}$ M$\odot$ clusters. These represent the some of largest halo-like structures supported at $z\sim0$, exceeding the mass of the Virgo\citep{2020A&A...635A.135K} and Coma\citep{2009A&A...498L..33G} Clusters. Further investigations into how our code handles these extremes will follow in future work.

\subsubsection{AGORA}
\label{sec: AGORA}

\begin{figure*}[t!]
\begin{center}
\includegraphics[width=\linewidth]{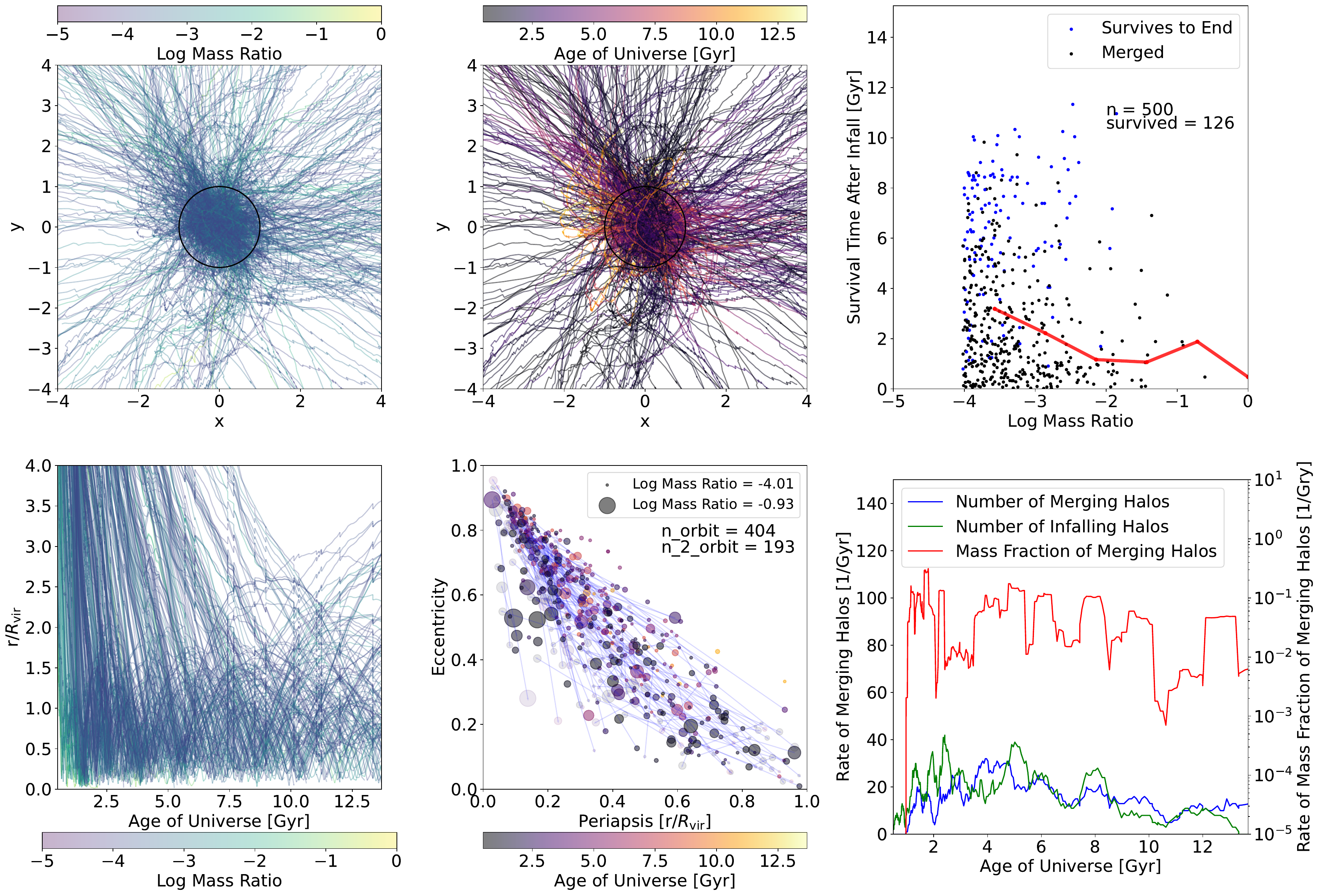}
\caption{Several diagnostics of the dynamics of the five hundred most massive halos that infall into the main halo in the AGORA's ART-I \texttt{Cosmorun-2} simulation using data from our halo-finding methods. Top left: A projection of halo dynamical tracks projected onto the simulation x and z plane and colored by the maximum mass ratio of the halo. Top center: Same as the top left but colored by cosmic time along the track (where longer-lasting halos show more of the color gradient along their path). The black circle represents the main halo radius ($r_{200}$) in both the top left and top center plots. Bottom left: Evolution of distance for infalling halos as a function of cosmic time and colored by mass ratio. Bottom center: The evolution of the eccentricity-periapsis relation of infalling halos colored by cosmic time and scaled by halo mass ratio. The first (solid) and second (translucent) orbital parameters as defined in Sec. \ref{sec: infall} for a halo are connected with a line. Top right: The survival time of infalling halos after first infall plotted against mass ratio with a bin average line plotted in red. Halos that are still present at the end of the simulation at $z=0$ are colored blue and halos that are not are colored black. Bottom right: Infall rate (green), and merger rate (blue) plotted on the right y-axis in per Gyr as well as the mass inflow rate from infalling halos (red, right axis) plotted against cosmic time.}
\label{fig:ART Shinbad}
\end{center}
\end{figure*}

Though we have identified some theoretical evidence to support halo suppression at the low mass range for our N-body results, we need perform a comparison to results from other halo finders to understand our results in the context of existing theory. Note that we do not compare the simulation codes, but rather use AGORA data to test and generalize our halo-finding algorithms.

In our analyses of the ARGOA simulations, RCT and {\sc Haskap Pie} were run and then limited to the refined region (as defined in Sec. \ref{sec:Refined Regions}), which we adjusted 11 times in concert with the overdensity-finding step to ensure that halos found in that step were limited to refined particles. We use a set of parameters for RCT that were painstakingly developed and tuned specifically to return and track a large number of satellites around the main halos in these data sets \citep{2024ApJ...964..123J}. We also do not enforce any additional restrictions on minimum halo mass on results for either finder or any other cut-offs in our analyses.  Results for RCT should therefore represent expert use of the code and a fair basis for comparison for any improvements or deficiencies in our algorithms. Note that the `RCT' results we compare to in this and the following sections refer to the results from {\sc Consistent Trees} made after running it on halo lists from {\sc Rockstar}. Also note that {\sc Consistent Trees} has routines that modify the halo list in order to connect trees in some cases so the comparisons we make are not directly to {\sc Rockstar} itself.

\begin{figure*}[t]
\begin{center}
\includegraphics[width=\linewidth]{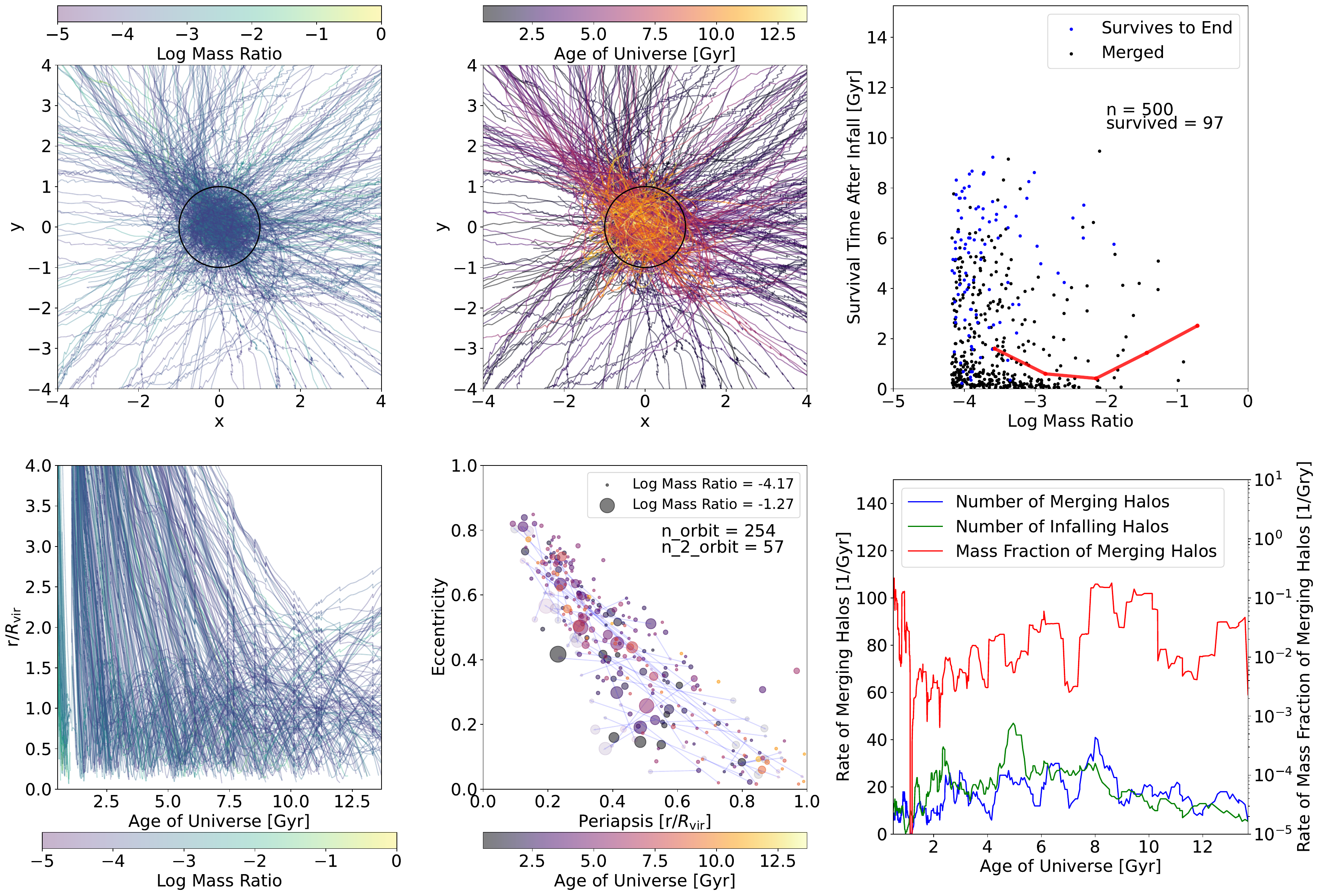}
\caption{Same as Fig. \ref{fig:ART Shinbad} compiled with RCT halo-finding results.}
\label{fig:ART RCT}
\end{center}
\end{figure*}

For our zoom-in regions, we do not expect our halo mass functions to conform to linear theory, especially at the high mass end since the simulations were refocused on rare peaks of overdensity to concentrate computational resources on a Milky Way-mass progenitor. Data for each simulation is available to $z=2$, which is approximately the redshift where halo counts peak. As shown in Fig. \ref{fig:AGORA HMF} (top row, left three plots), halo populations in each of the simulation runs approach linear theory for $5 \geq z\geq2$ for halo masses between a few 10$^7$ M$_\odot$ (a few hundred particles) and  10$^{11}$ M$_\odot$ for both RCT and {\sc Haskap Pie}. Below this value, both finders diverge from linear theory significantly with our halo-finder falling faster, which we further examine throughout our analysis of AGORA's ART-I halo-finding results. However, when there are enough particles for halos to be well defined regardless of method (M$_h$ $>$ 10$^8$ M$_\odot$), our halo-finder consistently detects and tracks more halos than RCT with halo number densities ranging from  20\% to 200\% higher. At the highest mass end (M$_h$ $>$ 10$^{11}$ M$_\odot$) results are stochastic but continue to generally show more halos using our method.

In Fig. \ref{fig:AGORA HMF} (bottom two rows), we show the ratio of the halo mass function to linear theory for each simulation, showing more redshifts for simulations that ran longer. We see that for $z \leq 5$, lower redshifts correspond to lower ratios to linear theory for all simulation suits and both halo-finders, which is consistent with our results for the N-body simulation in Sec. \ref{sec: n-body}. At all redshifts $z \leq 5$ RCT tends to over-perform our results for M$_h \lesssim$ $5 \times 10^{7}$ M$_\odot$ and underperform above $10^{8}$ M$_\odot$.

%Below $\sim100$ particles, an energy-based definition of a halo is less consistently defined, especially within complex potentials. 

\subsection{AGORA's ART-I Halos and Subhalos}

To understand the lower suppression of smaller halos in RCT, we embark in a more detailed study of the sub-halo population and merger behavior to distinguish the effects of particle-tracking on the survivability of subhalos and the nature of the low mass halos that might be missing from our results. 

Dedicated sub-halo finders have been shown to extend the halo mass function distribution to lower masses. Recent work by \citet{2025arXiv250206932F} showed that the choice FoF, bound-mass tracking, and hierarchical particle assignment routine can affect the resulting sub-halo population. Additionally, RCT can report a great number of small halos if one bypasses RCT's pruning routines with a small force resolution. The equivalent choice in our algorithm is to remove the minimum particle number restriction or to remove our own pruning procedure. However, it is prudent to dissect the halo population that results from each algorithm's best efforts to more deeply understand the differences before potentially over-fitting to a desired distribution.

Since the AGORA simulations are focused on around a single $10^{12}$ M$_\odot$ halo and its Lagrangian region, the mergers and subhalos of the main halo are the richest concentration of small halos in the refined region. We focus our evaluation of small halo tracking on data retrieved from the \texttt{Cosmorun-2}'s ART-I simulation, which was run to $z=0$. Fast data reading and an efficiently sized refined region made creating trees for this simulation faster than for our other AGORA simulations. Therefore, we were best able to iterate and refine our halo-finding parameters in response to results from ART-I. In our comparisons, we found that despite this tuning, results and analyses for ART-I are representative of all eight codes, with only minor differences in the number and timing of halos that interact with the main halo. This work will focus on the relative performance of our technique in various contexts, whereas a detailed analysis of merger dynamics in the entire AGORA suite including halo-finding results from {\sc Haskap Pie} and simulation code comparisons will be examined in forthcoming work by the AGORA Collaboration (Nguyễn et al., in prep.).

\subsubsection{Halos Interacting with the Main Halo}

The main halo, which acts as the source of the Lagrangian refined region, has a bound mass of $\sim 1.05 \times 10^{12}$ M$_\odot$ at $z=0$ to $\Delta_c =200$. To study halos that merge or make close passes to this halo, we apply an 3rd-degree Savitzky-Golay filter with a window size of 11 to the halo positions to lightly smooth halo tracks in both RCT and {\sc Haskap Pie}. Then we applied the following restrictions on our sample:

\begin{enumerate}
    \item Halos must persist for at least five timesteps within a box centered on the main halo and extending outward for four radii in each Cartesian direction to be included.
    \item Halos must have a maximum mass at any time step of at least $10^{7}$ M$_\odot$ (35 dark matter particles).
    \item Halo timesteps must be consecutive.
\end{enumerate}

These restrictions limit the number of spurious halos in either code and match the pruning condition of our code, where halos with less than five timesteps in their tracks are completely removed from our halo trees. The second condition avoids the contamination of resolution effects by keeping particle counts high enough to give both halo-finders are reasonable chance at completeness. This condition partially favors RCT since the minimum halo size was set to 35 dark matter particles for {\sc Rockstar} and our halo finder requires at least 11 particles in the energy-solving step and 6 particles for particle-tracking. However, results from RCT nonetheless include two-particle halos. This occurs because the virial mass reported by {\sc Rockstar} will usually be a subset of the mass of the corresponding particle limit as the limit only applies to the particles that are assigned to the halo, not the particles that comprise the virial mass. When the third restriction is relaxed, sharp straight halo paths across time are produced as RCT associates halos across non-consecutive timesteps. %(see Sup Fig. \ref{fig:ART RCT relaxed}, top left and center), especially at high redshift. Though in Sup Fig. \ref{fig:ART RCT relaxed}, we have relaxed this restriction to show these outliers and the complete sample, we report numbers for the restricted sample in our analyses. 
The underlying cause of these connections is not immediately clear, but they are inaccurate and unphysical as well as absent from results from {\sc Haskap Pie}. They could be due to errors in the gravity-based trajectory solving technique in {\sc Consistent Trees}. Results for RCT are generally better for halo survival times and rates, for example, in the restricted sample than when we relax the consecutive timestep condition so the restrictions do not generally disadvantage RCT in our comparisons. %Plots of our entire sample with these restrictions applied are available in Sup. Figs. \ref{fig:ART shin full} and \ref{fig:ART RCT full}. 
Note that these restrictions do not apply to the halo mass functions in Fig. \ref{fig:AGORA HMF}.

After the restrictions are applied, the remaining halos are then divided into two categories. The first category is halos that have minimum distances between their center of energy and the center of energy of the main halo of less than one halo radii and initial distances of greater than 1.5 halo radii, have undergone ``infall". At infall (time of earliest crossing of the main halo radius), their mass, velocity, and position are recorded. Though both halo trees extend to high redshift, we focus on halos that have infall times after 500 Myr as a larger fraction of halos is affected by resolution effects and the second restriction at earlier times. The second category is halos that are never closer to the main halo than one halo radii and are categorized as local, non-infalling halos. These categorical definitions leave a third category of excluded halos, which we discuss in Section \ref{sec: Excluded}.

\subsubsection{Infalling Halo Orbits}

\label{sec: infall}

Paths of the five hundred most massive infalling halos for our algorithm are displayed in Fig. \ref{fig:ART Shinbad} and the path of the five hundred most massive infalling halos in RCT are shown in Fig. \ref{fig:ART RCT}. The top left and top center plots of both figures show the x-y projected paths of infalling halos within 4 $r_{200c}$ colored by the mass ratio of the infalling halo to the main halo at the time of infall and the age of the universe, respectively. While both components of a major merger preserve their dense cores long into their interaction, the main halo definitions are altered by their overlapping potentials in both methods, which can be seen as discontinuities in the paths of infalling halos in the bottom left plots of Figs. \ref{fig:ART Shinbad} and \ref{fig:ART RCT} when the halo center and/or halo radius abruptly changes. In both halo-finding schemes, this is due to our allowance that particles outside of halo cores can belong to more than one halo.

Distinguishing halos during close passes to the halo center is a particularly challenging problem to solve for several reasons. Any spherical overdensity drawn about overlapping centers will mostly consist of members of the larger halo and return an overdensity-based halo radius equivalent to the main halo radius. Bound particle tracking will also struggle to discern between overlapping potential wells and segregate members. Halo mass loss due to dynamical friction will tend to spread sub-halo constituents throughout the main halo, making it difficult to recover useful position information. Additionally, since FoF-based algorithms return hierarchical assemblies of halos that may be overlapping definitions of the same halos, procedures to remove these overlapping halos will usually have a higher false-positive rate when the centers are closest to being coincident. The consequence of being unable to track close passes is a broken halo track and loss of the halo from the tree.

A key advantage of our halo-finder is that our halos are tracked much closer to the center of the main halo as shown in the bottom left plots of both figures. This is achieved by a combination of tracking bound particles, our multi-step halo finding technique, and our use of the center of bound mass velocity as a criterion in our pruning algorithm. The closest approach to the center of the main halo in the ARGORA ART simulation analyzed with our algroithm is about 1.12\% of the halo radius while the closest approach with RCT is about 3.86\% of the halo radius. Furthermore, in the 500 most massive halos of the {\sc Haskap Pie} results, 34 halos pass more closely to the center of the main halo than the closest pass of any of the the 500 most massive halos identified by RCT.

To determine whether halos are being disrupted near the halo center, we also compare the minimum periapses of both codes for the 500 most massive halos. In the center-bottom plots of Fig. \ref{fig:ART Shinbad} and Fig. \ref{fig:ART RCT}, we show orbit parameters for infalling halos that have an initial periapsis, $r_p$, inside of the radius of the halo, retreat to an apoapsis, $r_a$, and then fall closer to the halo center, thus completing most of an orbit. As shown in these plots, the closest periapses were about 2.77\% of the halo radii for {\sc Haskap Pie} and 9.14\% for RCT, which greatly exceeds the difference in the definition halo center. Therefore, we can conclude that that many more halo tracks are affected or interrupted near the main halo center in RCT.

In the center-bottom plots of Fig. \ref{fig:ART Shinbad} and Fig. \ref{fig:ART RCT}, the eccentricity of the orbit is calculated as $e = (r_a-r_p)/(r_a+r_p)$ and not from the eccentricity vector or angular momentum, which we explore in Sec. \ref{sec: ang mom}. The size of the points in the scatter plot represents the mass ratio of the infalling halo and their color represents the age of the universe at infall. In both results, we see that most halo orbits lie near a linearly inverse relationship between relative periapses ($r/r_{\rm vir}$) and eccentricity. This relationship implies that low eccentricity, low periapsis orbits (close-in circular orbits), and high eccentricity, high periapsis (radial orbits that miss the center) are disfavored. This is in line with the expectation that the trajectory of halos infalling from outside the main halo is not circularized at a close radius as dynamic friction will continue to bleed orbital energy. This is also in line with the expectation that radial orbits would tend to target the main halo center of mass. Low eccentricity orbits ($e < 0.2$) are established down to half the halo radis, however, which suggests that temporary circularization does occur at larger radii.

There are far fewer halos with established orbits (defined as having a periapsis, apoapsis, and second periapsis after infall) in RCT results than with {\sc Haskap Pie}. Overall, 80.8\% of the 500 most massive halos establish an orbit in our algorithm and 50.8\% in RCT. This implies that for halos infalling into this main halo, RCT is $\sim 2.5$ times as likely to lose track of the halo within the first orbit. In the RCT results, periapses below $r/r_{\rm vir} \sim 0.15$ are largely missing with a single high eccentricity exception in the unpruned results, but there are still far fewer established orbits with $r/r_{\rm vir} > 0.2$, which implies that halo-tracking is failing for reasons in addition to issues related to tracking close halo passes. 

Second orbits (parameters based on the second periapsis and second apoapses if they exist) are displayed in the center bottom plots with translucent markers connected to their first orbit parameters by a translucent line. Second orbits are established for 38.6\% of halos in {\sc Haskap Pie} as compared to 11.4\% of halos with RCT, implying that RCT is 48.6\% more likely to lose track of a second orbit after establishing a first orbit (77.6 \% lost versus 52.2\% lost). The loss of second orbits can be partially explained by timing. As shown in the bottom left plots of the figures, there is a strong relationship between the period and semi-major axis of orbits and redshift. For a halo orbit with a semi-major axis equal to the radius of the main halo, using Kepler's Second Law, the orbital period becomes 
\begin{equation}
    T = t_{\mathrm{ff}}\times4\sqrt{5.38/\Delta_c},
\end{equation}
where $t_{ff}$ is defined in Eq. \ref{eq: tff} and is not a function of main halo mass. The current day value of $T$ is approximately 9.133 Gyr and the value at $z=10$ was approximately 250 Myr. Therefore, many of the later-infalling halos have not had an opportunity to make a second orbit. However, as discussed in Sec. \ref{sec: survive} and shown in the top right plots, far fewer halos make it to the end of the simulation than establish a first orbit and so both {\sc Haskap Pie} and RCT lose track of halos after their first orbit. 

As a secondary effect, the lengthening orbital periods also delay the dissipation of halos. This delay time means that the merger rate tends to lag the infall rate and is thus generally higher after the infall rate peaks. %(bottom right plots in Supplemental Figs. \ref{fig:ART shin full} and \ref{fig:ART RCT full}).
Since major mergers bring their own halo complexes, the overall infall and merger rate of the main halo is often dominated by halos that come during these events rather than the suppression of the overall halo mass function. The mass inflow rate from mergers (bottom right, red line) is far more uneven than the rate infalling and merging of halos and shows no favored epoch or peak when plotted as a mass ratio to the main halo.

When a halo-finder is more likely to lose track of halos somewhere between the first and second orbits, the distribution of early infalling to late infalling halos should be roughly consistent with a peak at $z \sim 2$ as discussed in our N-body analysis (Sec. \ref{sec: n-body}) and the halo tracks in the top center plot should be biased towards later times. This effect is more prominent in the RCT results, which show warmer colors (later times) for tracked orbits in the top-center plots of both figures. %While we do expect that, dark matter halos can become dissipated, dispersed, or intermixed to the point that they lack a physical definition.
The discrepancy with RCT implies that loss of tracking is not strictly due to these processes and that our halo finder is more robust for applications where subhalo orbits need to be tracked.

\subsubsection{Excluded Halos}
\label{sec: Excluded}

\begin{figure*}[t!]
\begin{center}
\includegraphics[width=\linewidth]{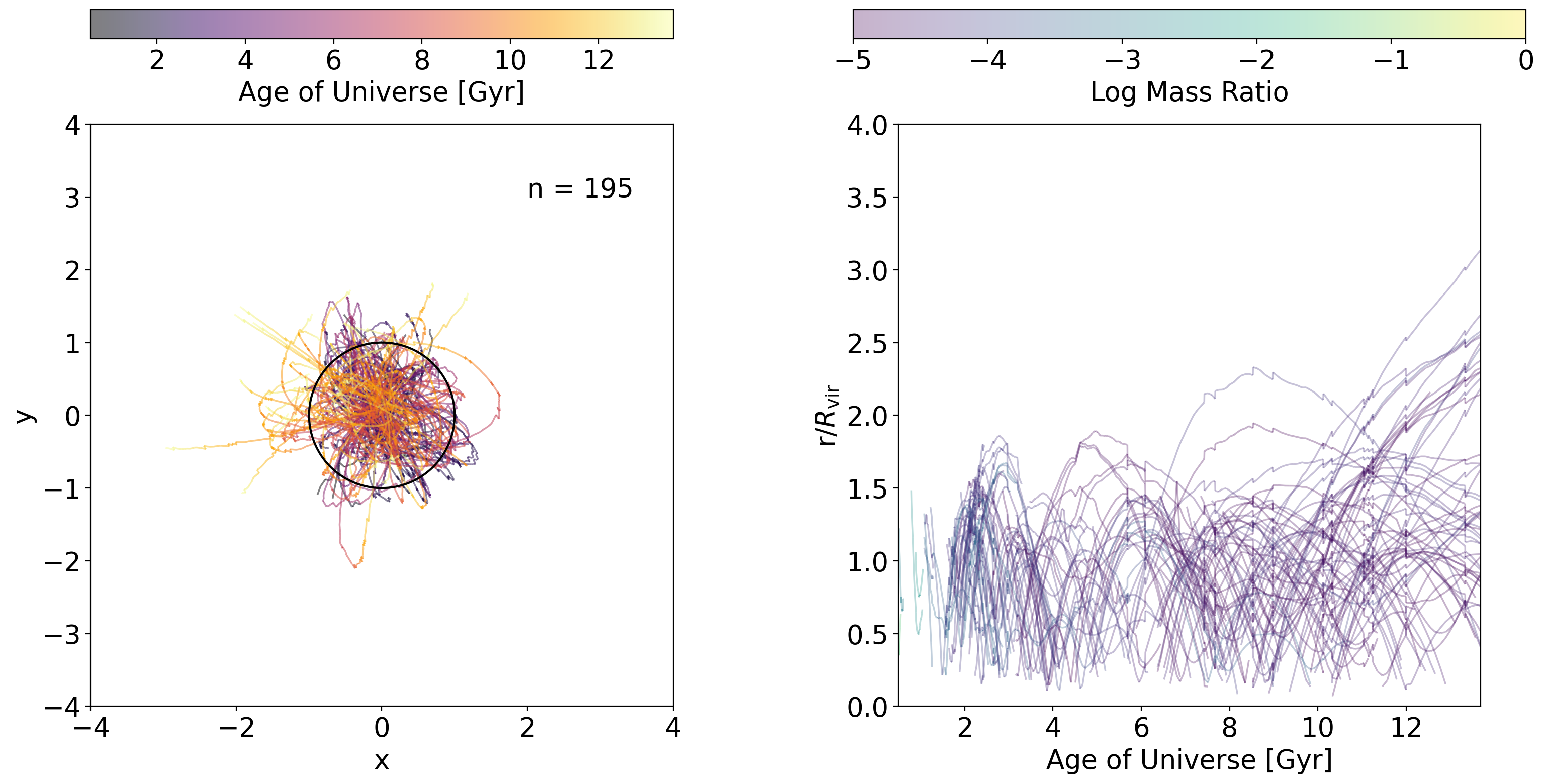}
\caption{Halos excluded from the RCT sample due to having earliest locations in their history within 1.5$r_{\rm vir}$ of the main halo,  implying that RCT lost track of their connectedness to halos earlier during the assembly of the main galaxy's complex.  Left: The paths of the halos that were excluded colored by the age of the universe. Right: The radial evolution of the excluded halos colored by their maximum mass ratios. Both plots confirm that the halo tracks are broken and incomplete for at least 195 halos for of the top 500 halos that were not excluded. The corresponding plot for our halo finder was not included because only eight halos failed the same test with our method. }
\label{fig:ART Excluded}
\end{center}
\end{figure*}

Halos that begin closer than 1.5$r_{\rm vir}$ away from the main halo but also have their minimum distances to the main halo smaller than $r_{\rm vir}$ are excluded from the sample. This applies to eight halos for our halo finder and 195 halos for RCT. The path of excluded halos in RCT is plotted in Fig. \ref{fig:ART Excluded}. The halos excluded in this manner are heavily biased towards larger halos as the excluded halos were all in the top 500 halos by mass and are likely double counted as branches in the halo tree. This also highlights the discrepancy between halo lists, which may show a large number of halos at any given time, and halo trees, which only show halos with progenitors or descendants at other times. We again emphasize that without temporal tracking, halo lists can easily include non-halo clusters of particles that cannot be confirmed or studied dynamically.

Excluded halos in RCT are also biased towards halos with large, looping orbits as these are easiest to track when they are most distant from the main halos. In our halo finder, all excluded eight halos are short, non-Keplerian paths within the main halo. We were able to roughly match 110 of the excluded halos in the RCT results to results from our finder by determining the position and timing when halos in RCT are at 1.75 $\times r_{\rm vir}$ away from the main halo and then searching {\sc Haskap Pie} halo lists for halos that were rough matches in position, radius, and velocity. Of this sample, our corresponding halo tracks were 4.7 Gry longer on average. At least 45 of those 110 halo tracks were double counted as separate RCT tracks along a single track reported by our finder.

We have not excluded halos that are tracked to beyond 1.5$r_{\rm vir}$ but may be the same physical halo that is counted separately during another pass as there are no easy ways to determine the connectedness of halos using RCT results. Therefore, RCT results may represent an undercount of broken-track halos. The relative lack of broken halo track with our halo-finder can be regarded as a relative strength of our approach in tracking sub-halos, but is also a result of our pruning strategy.

\begin{figure*}[t]
\begin{center}
\includegraphics[width=0.8\linewidth]{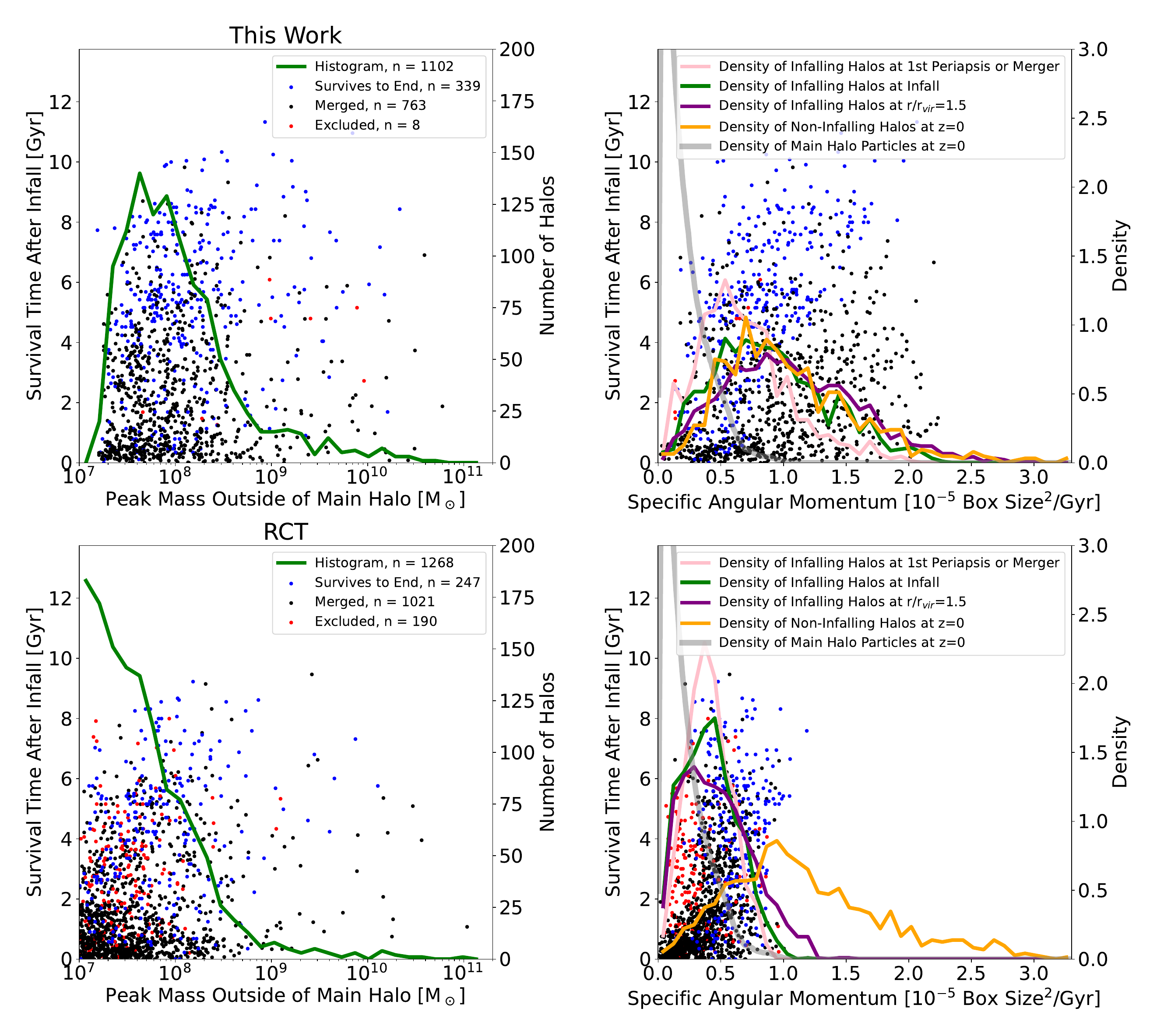}
\caption{Left column: Halo demographics for the full sample of halos falling into the main halo in the AGORA's ART-I simulation as a function of halo mass and colored by whether the halo survives to $z=0$ (blue), is lost to the halo-finder before $z=0$ (black) or is excluded from some of our analyses due to the halo track starting within the influence region of the main halo (r < 1.5r$_{200}$) (red) along with a bin histogram of peak mass outside of the main halo shown as a green line for our halo finder (top) and RCT (bottom). Right column: Angular momentum distribution with respect to the main halo for infalling halos. Bin histograms are also added for halos at first periapsis (salmon), at first infall (r=r$_{200}$) (green), infalling halos at r=1.5r$_{200}$ before first infall (purple), the particle distribution at $z=0$ (gray), and halos that do not infall but are within 4r$_{200}$ at $z=0$ (orange). The top plots show results for {\sc Haskap Pie} and the bottom plots show results from RCT. The infalling halo angular momentum distribution is highly skewed toward the particle distribution for RCT and consistent with the halo population outside of the halo radius for our method. Many of the excess halos in the RCT sample are close to the mass resolution and have very low reported angular momentum.}
\label{fig:ART Momentum}
\end{center}
\end{figure*}

\subsubsection{Survival Times}
\label{sec: survive}

We define a halo's ``survival time" as the time it takes a halo-tracking algorithm to lose track of a halo after infall. On average, the five hundred most massive infalling halos meeting all three restrictions have survival times of 3.58 Gyr after infall in our model and 2.78 Gyr in RCT.  Furthermore, as shown in the top right plots in both Figs. \ref{fig:ART Shinbad} and \ref{fig:ART RCT}, 126/500 halos survive to $z=0$ after infall in our model and 97/500 do in RCT.  

Of the full sample of halos reported by both halo-finders that meet our conditions for inclusion, 339/1102 (30.7\%) survive to $z=0$ in {\sc Haskap Pie} and 247/1268 (19.5\%) in RCT. Survival times in our model are about 1.4 Gyr longer on average (6.859 Gyr versus 5.464 Gyr) for the full sample. While Fig. \ref{fig:AGORA HMF} already showed that there are more low-mass halos ($<$ 100 particles) reported by RCT, we find that our low-mass halos have longer survival times in our model as plotted in Fig. \ref{fig:ART Momentum} (left column). This means that the smaller halos in RCT are much shorter-lived (1.65 Gyr for RCT versus 2.78 Gyr for our model for halos with maximum masses before infall less than $10^{8}$ M$_\odot$). 

These survival times are all tremendously long, which challenges the definition of a merger and the notion of a merger rate. This is in keeping with the dynamics of collision-less dark matter, which is immune to baryonic stripping mechanisms. 

For the halos that infall with a high mass, often the interaction pulls both halos together and any orbits of the infalling halo are biased towards low periapses. This means that RCT results, which are challenged during close passes, are more likely to lose track of a major merger. Though our halo-finder also struggles in the scenario, we can more easily track halos that are self-bound with our method and so all the major mergers are tracked with {\sc Haskap Pie} are tracked beyond the first periapsis and in most cases, well after the halo is severely disrupted.

\subsubsection{Angular Momentum Discrepancies}
\label{sec: ang mom}

We investigated the dynamical properties of merging halos and subhalos to further probe the demographic differences between the halo populations that were calculated in {\sc Haskap Pie} and those found with RCT. Our algorithm has the benefit of using bound particles to track the movement of halos and tends to ignore untrackable dense concentrations of particles. One method to determine if a sub-halo is truly independent of the N-body particle cloud of a main halo is to examine the relative angular momentum distribution of halos with respect to the main halo.

In Fig. \ref{fig:ART Momentum} (column two), we consider the co-moving angular momentum of particles within the halo radius (gray) as well as for infalling halos at various distances and non-infalling halos. Using a co-moving unit allows us to compare values for infalling halos over cosmic time. Particles within the halo radius generally have a nearly-Poisson binomial distribution with a peak at very low angular momentum, which corresponds to radial orbits. Halos that are within the four halo radii, but that have not fallen within the radius of the main halo tend to have a broad angular momentum distribution. As merging and subhalos fall into the main halo, we expect some bleeding of angular momentum, especially in situations most affected by dynamic friction, but for it to be mostly conserved in the pre-infall state.

However, infalling halos reported by RCT tend to have angular momentum distributions much closer to the particle distribution and almost no infalling halos with higher angular momentum as shown in Fig. \ref{fig:ART Momentum} (bottom right), which is physically implausible. The halos ``excluded" from our merger analysis due to not having an origin outside of the main halo are shown in red for both methods, showing that there is a direct correlation between halo-tracking and angular momentum for RCT and no corresponding issue with {\sc Haskap Pie}. Because many of the excluded halos have low reported angular momentum, this apparent mixing between halo particles could be posing a challenge for the tree-building routines in {\sc Consistent Trees}, especially for lower mass halos. In general, the vast majority of the infalling halos that RCT tracks are short-lived, low mass, and are reported to have low angular momenta with respect to the main halo. Whereas our halo-finder (column two, top) shows a distribution exhibiting mild dynamical friction effects as halos fall inward, as expected. Since at least half the excluded halos are present and well-tracked by {\sc Haskap Pie} and are included in the top right plot of Fig. \ref{fig:ART Momentum}, matched halos show a clear discrepancy in angular momentum at the time of infall.

This reveals a stark difference between the results. Sub-halo and merging halo velocities reported by RCT are closely correlated to the particles of the main halo, which could provide a challenge to dynamical studies of these halos. Whereas our results for halos larger than about 100 times the dark matter mass resolution are both more complete and are more plausible dynamically since they maintain their velocity distribution as they cross the radius of the main halo. In both algorithms, halo velocities are used to track them through time and so an incorrect velocity hampers halo-tracking for RCT. This reveals that the quantity of infalling halos reported by the RCT analysis of the AGORA's ART-I simulation may not correspond to the efficacy of its halo-finding and tracking routine and that our smaller sample of infalling halos for this main halo is not strictly a subset of the RCT sample and may include halos that RCT cannot track.

\subsubsection{Other Halo Groups}
\label{sec: other groups}

\begin{figure}
\begin{center}
\includegraphics[width=\linewidth]{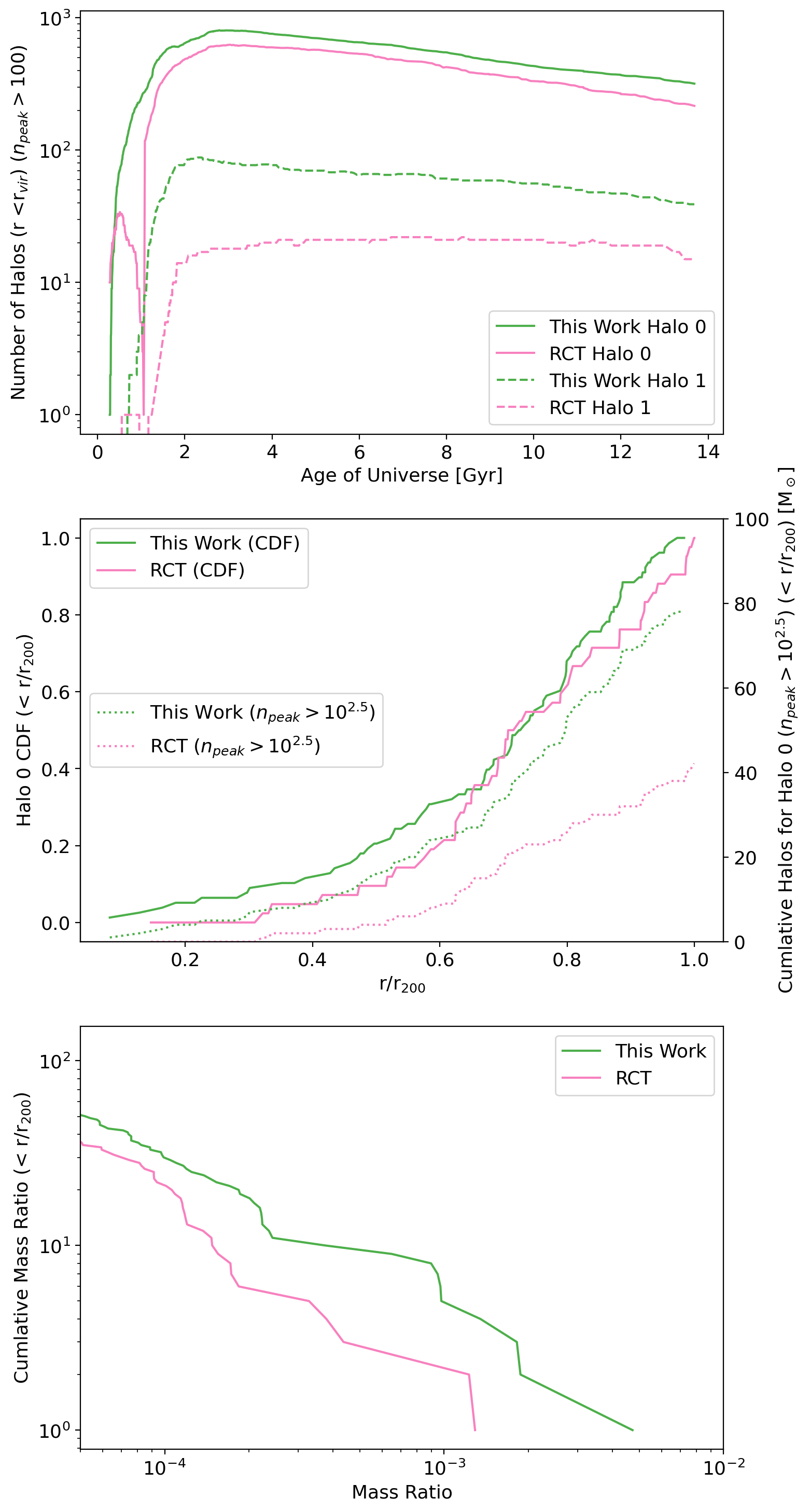}
\caption{Three measures of halo counts within r$_{200}$ of a main halo with comparisons between RCT and our method for the AGORA's ART-I simulation.  Top: Halo counts for halos within the largest and second-largest halos as a function of cosmic time. Only halos with $n_{peak}$ > 100 particles are included, where our solutions converge.  Center: Partial recreation of \citep{2024ApJ...970..178M} (Figure 11) showing the radial counts of halos at $z=0$.  The left axis and solid lines show the cumulative density function and more halos at smaller radii as a fraction of the total. The right axis and dotted lines show the total number of halos, showing far more in our method for $n_{peak}$ > 10$^{2.5}$ (to match the source plot). Bottom: Partial recreation of \citet{2024MNRAS.533.3811D} (Figure 7) showing cumulative mass ratios of the halo population within r$_{200}$ showing a boost over RCT throughout the range of mass ratios.}
\label{fig:peak_part}
\end{center}
\end{figure}

Halos other than the target halo were also solved with both algorithms, which allowed us to place our comparisons in context as well as begin to generalize them. Trees for the second most massive halos were far more dense for {\sc Haskap Pie} than for RCT, featuring about two and a half times as many infalling halo tracks (112 versus 45), tracking ten times as many orbits (72 versus 7), and three times as many halos survive until the last timestep (39 versus 13).  Fig. \ref{fig:peak_part} (top), shows the number of halos within the radius of the largest and second-largest main halos in the AGORA's ART-I simulation.  Here we enforce that the peak mass of the halo, $n_{peak}$, is at least 100 hundred times the dark matter mass resolution.  When we limit the sample in this way, our halo mass functions for our algorithm were generally more dense than for RCT with the effect becoming more pronounced for smaller main halos.  Even when including all halos regardless of particle number, our algorithm returns more halos at $z=0$ for both halo groups as they are more likely to be tracked through to the end of the simulation after their infall. 

\subsection{Qualitative Comparisons with Other Codes}

Our comparisons with RCT were supported by robust data generated from a concerted effort to tune RCT's parameters within the AGORA collaboration to study satellite galaxies, resulting in many more and better-tracked subhalos than RCT's default settings. This expert use of {\sc Rockstar} allowed us to provide a high baseline for the results of our halo-finder. Several other halo finding and tracking algorithms have been developed that extend or improve on RCT, and not all versions were readily available to use for a like-to-like comparison with our results, nor do we have access to data derived from similarly developed expertise. Therefore, we have made a pair of qualitative comparisons between our results to data from their methodology papers. 

\begin{figure*}[t]
\begin{center}
\includegraphics[width=\linewidth]{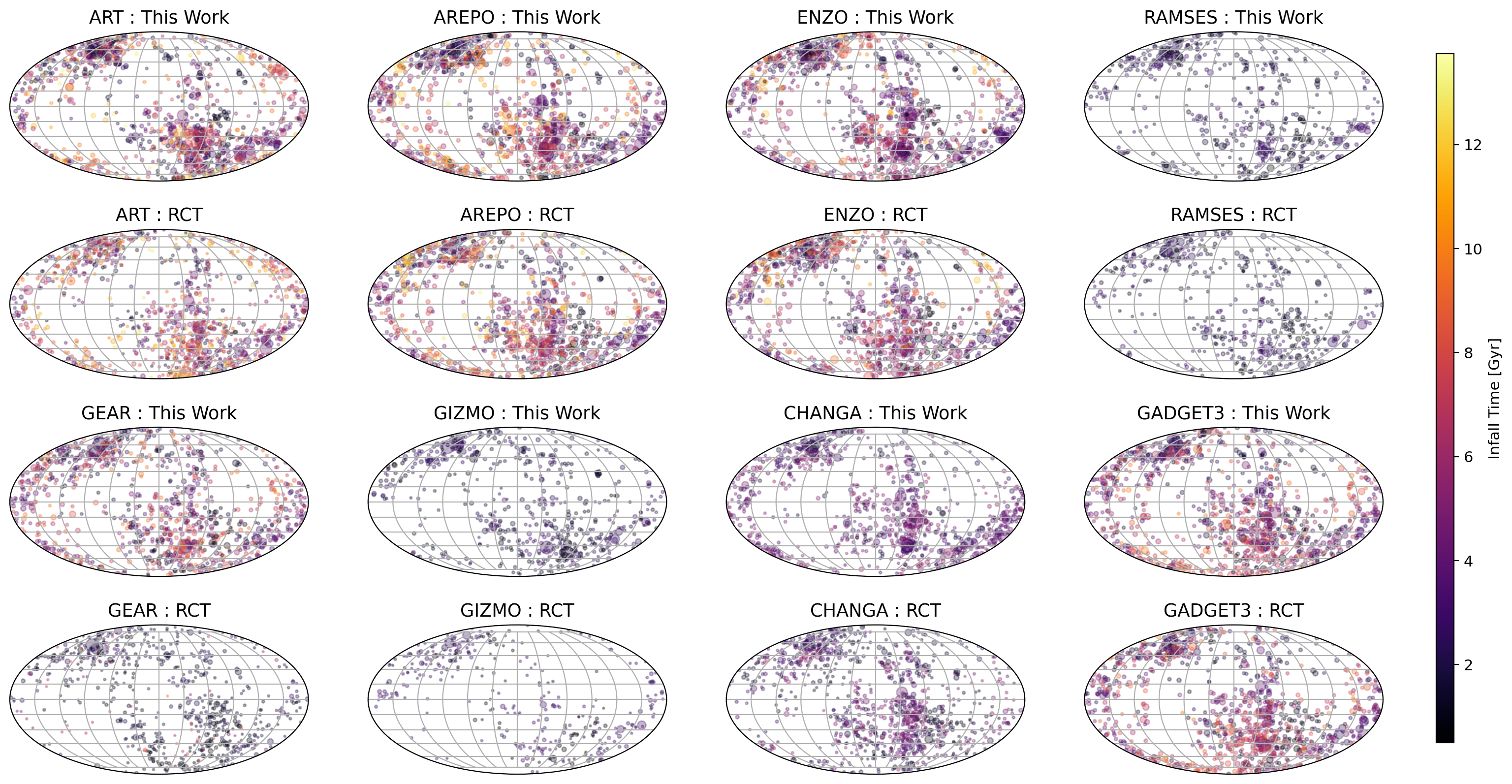}
\caption{Mollweide projections of the first infall point of the 500 most massive infalling halos onto the radius of the main halo plotted for all eight AGORA simulations to their final available redshift and colored by the time of infall for both RCT and the results of this work. The polar/vertical axis is aligned to the $z$ coordinate of each simulation. All simulations show the same, mostly time-independent, concentrations of infall points with the exception of GEAR and GIZMO for which RCT data is incomplete and we needed to recreate the particle data in order to complete our results.  The clustering is suggestive of a filamentary structure that does not fundamentally shift with time about the main halo. We also see that the five hundred most massive halos are generally larger than the five hundred most massive halos found with RCT by the scatter marker sizes, which are scaled with peak infall mass ratio. A detailed analysis on merger dynamics in the AGORA suite and simulation code comparisons will be examined in forthcoming work by the AGORA Collaboration (Nguyễn et al., in prep.).}
\label{fig:Molly All}
\end{center}
\end{figure*}

\subsubsection{\sc{Symfind}}

The method of {\sc Symfind} \citep{2024ApJ...970..178M} focuses on extending the tracks of halos found with RCT to address issues tracking halos through infall and more robustly tracking halos as they experience mass loss and disruption. The code greatly extends halo tracks and similarly finds that {\sc Rockstar} can struggle to track halos at periapsis.

We reproduce part of their Figure 11 showing the radial cumulative distribution of halos in their simulation suite using data from the main halo at $z=0$ from our AGORA's ART-I analysis. Their figure compares halos that, at their peak, $n_{\rm peak}$, contain more than $10^{2.5}$ particles so we make the same cut for our comparison. As shown in Fig. \ref{fig:peak_part} (center), our results show a larger fraction of our halos are closer to the main halo's center which compares to {\sc Symfind}'s results showing slightly more halos at more intermediate distances for the comparable sample. However, because our results are not tied to RCT like {\sc Symfind} and our finder tends to track higher-mass halos more consistently after infall, our cumulative number of halos is roughly twice the number as RCT at all radii for $n_{\rm peak} > 10^{2.5}$.

\subsubsection{\citet{2024MNRAS.533.3811D}}

\citet{2024MNRAS.533.3811D} similarly use RCT to identify candidates and focus on identifying and tracking subhalos that were lost with RCT. We partially reproduce their Figure 7 showing the halo mass function for sub-halos as a function of their mass ratio for the AGORA's ART-I main halo. As shown in Fig. \ref{fig:peak_part} (bottom), our results show a significant increase in the halo mass function for this main halo and show somewhat more of a boost as compared to the halo ranges presented in \citet{2024MNRAS.533.3811D} Figure 7. It should be noted that while their results were for a statistical sample and ours is only for a single halo, as discussed in Sec. \ref{sec: other groups}, our results overperformed for other halos in our simulations as well.

\subsection{Plane of Mergers}

Evidence of patterns or biases in the direction halos are infalling into the main galaxy were also examined and are presented in Fig. \ref{fig:Molly All} for both our halo-finding results and RCT for the 500 most massive infalling halos in all AGORA data sets. This number was chosen since results from all eight simulations had at least 500 infalling halos in our halo-finding results and so demographic differences between the samples could be studied. 

All results show that halos are infalling primarily from two oppositional regions over all cosmic time, which is consistent with the presence of filaments. However, our halo-finding results show more subhalos halos joining with larger infalling halos, which appears in the figures as higher clustering of similarly colored markers (similar infall times) around more massive infalling halos (which are represented as larger markers). This implies that larger subhalos that are lost in RCT are retained in our routine, which is consistent with our analysis of the ART-I data.

For GEAR and GIZMO, RCT results are further hampered by issues with particle reading that seem to be similar to the effects we saw using \texttt{yt}. This issue was resolved with our particle-saving workaround. This effect does not manifest as a significant reduction in the total number of halos as compared to results from the other code but does result in a reduction in the number of halos that can be tracked to infall into the main halo, which appears as far fewer halos in Fig. \ref{fig:Molly All} (GIZMO : RCT), which is the only plot to have fewer than 500 samples ($n = 354$), and smaller halos especially at low redshift in (GEAR : RCT) as in both cases, larger halos are more likely to be affected by particle reading errors.

\section{Summary}
\label{sec: summary}

We have developed {\sc Haskap Pie}, a new stand-alone halo finding and tracking algorithm combining overdensity-finding, energy-solving, cluster-finding, and particle tracking into a robust solver that is versatile enough to be used on several simulation types across eight simulation codes (ART-I, ENZO, RAMSES, CHANGA, GADGET-
3, GEAR, AREPO, and GIZMO). Our algorithm is Python-based and easily combined with the \texttt{yt} suite of analysis packages, which makes it accessible to a large portion of the community. Despite being Python-based, our code is well-optimized and can solve science-scale halo trees on a personal computer or laptop or on a computing cluster with good scaling ($<O(n_o)$) at high particle counts.

During our testing and analysis, we have identified these key strengths of our methods as compared to the state-of-the-art:

\begin{itemize}
    \item We track infalling halos/subhalos for a significantly longer time than RCT.
    \item We track infalling halos/subhalos much closer to the main halo center than RCT.
    \item We recover more halos with over 100 particles than RCT.
    \item Our halo velocities and angular momenta are more physically consistent than RCT.
    \item Halo tracks are far less likely to be broken or inconsistent than RCT.
    \item We improve the quantity of halos we track as compared to RCT more than work presented by \citet{2024MNRAS.533.3811D} and \citet{2024ApJ...970..178M}.
    
\end{itemize}

In addition to being useful for generalized studies of halos and galaxies in cosmological simulations, our code is especially useful for studies of mergers and satellite galaxies, including forthcoming work by the AGORA collaboration (Nguyễn et al., in prep.). 

\subsection{Limitations and Future Work}

We have much work to do to optimize our code to run on extremely large simulations with many tens or hundreds of thousands of halos and many hundreds of saved timesteps. Our halo finder is less likely to track halos that are less than 100 particles than some other finders, including RCT, and should not be considered complete in that situation. We are also working to improve our definition of a ``merged'' halo so that it can be robust against outliers and form a complete picture of mergers. Finally, since we do not explicitly use a spherical overdensity definition of a halo, we have not included analyses of the typical parameters that assume spherical symmetry in our comparisons such as circular velocity and virial radius. A much more detailed investigation into our halo morphologies will follow in future work.

\section*{ACKNOWLEDGMENTS}

We acknowledge the AGORA collaboration for their support and simulation data products. KSSB, THN, and ECS acknowledge the University of Illinois at Urbana-Champaign for their continued support. KSSB acknowledges the National Center for Supercomputing Applications as well as the Texas Advanced Supercomputing Center for their support with the Delta Supercomputer and Stampede3 Supercomputer, respectively, as well as the ACCESS program for computing grants PHY230100 and PHYS240175. THN acknowledges support from the National Center for Supercomputing Applications Center and its Center for Astrophysical Surveys.  

\section*{Data Availability}

{\sc Haskap Pie} and associated AGORA halo trees will be made available to the public with primers and data-reading guides after internal and external review.

\bibliography{main}
\bibliographystyle{aasjournal}

% \section{Supplemental Figures}

% \begin{figure*}[t]
% \begin{center}
% \includegraphics[width=\linewidth]{path_plot_1268-0-2000-0.pdf}
% \caption{Same as Fig. \ref{fig:ART Shinbad} showing all infalling halos instead of the top 500 for {\sc Haskap Pie}.}
% \label{fig:ART shin full}
% \end{center}
% \end{figure*}

% \begin{figure*}[t!]
% \begin{center}
% \includegraphics[width=\linewidth]{path_plot_1268-0-2000-1.pdf}
% \caption{Same as Fig. \ref{fig:ART RCT} showing all infalling halos instead of the top 500 for RCT.}
% \label{fig:ART RCT full}
% \end{center}
% \end{figure*}

% \begin{figure*}[t]
% \begin{center}
% \includegraphics[width=\linewidth]{path_plot_1268-0-500-1_2.pdf}
% \caption{Same as Fig. \ref{fig:ART RCT} for RCT except that the condition that timesteps for a halo are consecutive is relaxed.}
% \label{fig:ART RCT relaxed}
% \end{center}
% \end{figure*}

\end{document}